\documentclass[twocolumn,preprintnumbers,amsmath,amssymb]{revtex4}

\usepackage{graphicx}
\usepackage{dcolumn}
\usepackage{bm}
\usepackage{graphics}
\usepackage[usenames]{color}
\usepackage[toc,page]{appendix}
\usepackage{hyperref}

\hypersetup{
    bookmarks=true,         
    unicode=false,          
    pdftoolbar=true,        
    pdfmenubar=true,        
    pdffitwindow=false,     
    pdfstartview={FitH},    
    pdftitle={My title},    
    pdfauthor={Author},     
    pdfsubject={Subject},   
    pdfcreator={Creator},   
    pdfproducer={Producer}, 
    pdfkeywords={keywords}, 
    pdfnewwindow=true,      
    colorlinks=true,       
    linkcolor=red,          
    citecolor=blue,        
    filecolor=magenta,      
    urlcolor=green           
}
\newcommand {\apgt} {\ {\raise-.5ex\hbox{$\buildrel>\over\sim$}}\ }
\newcommand {\aplt} {\ {\raise-.5ex\hbox{$\buildrel<\over\sim$}}\ }
\begin{document}

\preprint{}

\title{Phenomenological Ginzburg-Landau-like theory for superconductivity in the cuprates}
\author{Sumilan Banerjee$^*$}
\author{ T. V. Ramakrishnan$^{*+}$}
\author{Chandan Dasgupta$^*$}
\affiliation{$^*$ Centre for Condensed Matter Theory, Department of Physics, Indian Institute of Science, Bangalore 560012, India\\
$^+$ Department of Physics, Banaras Hindu University, Varanasi 221005, India}

\begin{abstract}

We propose and develop here a phenomenological Ginzburg-Landau-like theory of cuprate high-temperature
superconductivity. The free energy of a cuprate superconductor is
expressed as a functional $\mathcal{F}$ of the complex spin-singlet pair amplitude 
$\psi_{ij} \equiv \psi_m =\Delta_m\exp(i\phi_m)$ where $i$ and $j$ are nearest-neighbor sites of the 
square planar Cu lattice in which the superconductivity is believed to primarily reside and $m$ labels the 
site located at the center of the bond between $i$ and $j$. The system is modeled as a weakly
coupled stack of such planes. We hypothesize a simple form, $\mathcal{F}[\Delta,\phi]=\sum_m (A\Delta_m^2+
(B/2) \Delta_m^4)+C\sum_{<mn>} \Delta_m\Delta_n \cos(\phi_m-\phi_n)$, for the functional, where $m$ and $n$ are 
nearest-neighbor sites on the bond-center lattice. This form is analogous to the original continuum Ginzburg-Landau free
energy functional; the coefficients $A$, $B$ and $C$ are determined from comparison with experiments. 
A combination of analytic approximations, numerical minimization and Monte Carlo simulations is used to work out a number of 
consequences of the proposed functional for specific choices of $A$, $B$ and $C$ as functions of hole density $x$ and 
temperature $T$. There can be a rapid crossover of $\langle\Delta_m\rangle$ from small to large values as $A$ changes sign 
from positive to negative on lowering $T$; this crossover temperatures $T_\mathrm{ms}(x)$ is identified with the observed 
pseudogap temperature $T^*(x)$. The thermodynamic superconducting phase-coherence transition occurs at a lower 
temperature $T_c(x)$, and describes superconductivity with $d$-wave symmetry for positive $C$.
The calculated $T_c(x)$ curve has the observed parabolic shape. The results for the superfluid density  $\rho_s(x,T)$, 
the local gap magnitude $\langle\Delta_m\rangle$, the specific heat $C_v(x,T)$ (with and without a magnetic field) as well 
as vortex properties, all obtained using the proposed functional, are compared  successfully with experiments. We also 
obtain the electron spectral density as influenced by the coupling between the electrons and the correlation function of 
the pair amplitude calculated from the functional and compare 
the results successfully with the electronic spectrum measured through Angle Resolved Photoemission 
Spectroscopy (ARPES). For the specific heat, vortex structure and electron spectral density, only some of the final results 
are reported here; the details are presented in subsequent papers.
\end{abstract}
\maketitle

\section{Introduction}\label{sec.Introduction}
 
The last two decades have seen unprecedented experimental and theoretical activities involving cuprates which
exhibit high-temperature superconductivity \cite{PALee,KHBennemann,JRSchrieffer,MAKastner}. Even after
this long period of research which has seen dramatic advances in experimental techniques and
materials quality, as well as discovery of many unusual phenomena such as the ubiquitous pseudogap in underdoped cuprates
\cite{TTimusk,SHufner,MRNorman1,JLTallon} and the `strange metal' phase above the superconducting transition
temperature around optimal doping \cite{PALee,KHBennemann,JRSchrieffer}, there is no common, 
broadly accepted understanding yet about their origin.  

Motivated by the above, especially the increasing volume of sophisticated spectroscopic data on the cuprates
(such as those obtained from ARPES~\cite{ADamascelli,JCCampuzano}, STM~\cite{OFischer} and
Raman~\cite{TPDevereaux} experiments), 
we propose and develop here, as well as in subsequent papers, a new phenomenological
model for cuprate superconductivity that is analogous in form to the well-known Ginzburg-Landau (GL) theory 
\cite{VLGinzburg} of superconductivity. The starting point of our description is the assumption 
that the free energy of a cuprate superconductor can be expressed as a functional solely of the
complex pair amplitude. In the original continuum GL theory, the free energy, expressed as a functional of
the complex order parameter field $\psi(\mathbf{r})=\Delta(\mathbf{r})\exp{(i\phi(\mathbf{r}))}$, has the form 
\begin{eqnarray} 
\mathcal{F}(\{\psi({\bf r})\})&=&\int d{\bf r} \left[A_c |\psi({\bf r})|^2+\frac{B_c}{2}
|\psi({\bf r})|^4 \right. \nonumber \\
&+&\left. \frac{C_c}{2}|\boldsymbol{\nabla}\psi({\bf r})|^2 \right]. \label{Eq.continuumGLfunctional}
\end{eqnarray}
This form is justified near the actual superconducting transition where 
the magnitude of the order parameter is small, so that a 
low-order power series expansion in $\psi(\mathbf{r})$ is adequate. Further, $\psi(\mathbf{r})$ is assumed to vary slowly 
with $\mathbf{r}$ so that it suffices to keep only the $|\boldsymbol{\nabla}\psi(\mathbf{r})|^2$ term; this is the case in
conventional superconductors where the natural superconducting length scale (also the coarse graining scale)
$\xi_0$ is large (compared, say, to the Fermi wavelength). After the advent of the microscopic 
Bardeen-Cooper-Schrieffer (BCS) \cite{JBardeen} theory of
superconductivity, $\psi(\mathbf{r})$ was identified by Gor'kov \cite{LPGorkov} with the Cooper pair amplitude, 
i.e. $\psi({\bf r})=\langle a_{\uparrow}({\bf r})a_{\downarrow}({\bf r})\rangle$, where 
$a_\sigma(\mathbf{r})$ ($a_\sigma^\dagger(\mathbf{r})$) is the operator which destroys (creates) an electron
at $\mathbf{r}$ with spin $\sigma$ ($\sigma=\uparrow,\downarrow$). Gor'kov also obtained the coefficients
$A_c$, $B_c$, $C_c$ in terms of the electronic parameters of the metal.

\begin{figure}
\begin{center}
\includegraphics[height=6cm]{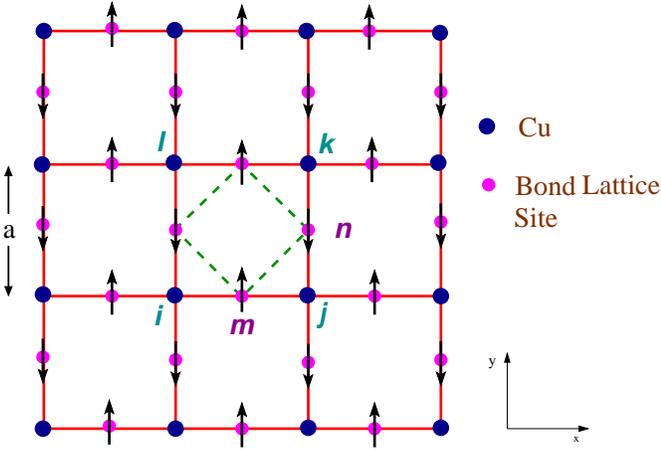}
\end{center}
\caption{The square Cu lattice sites $i,j,k,l,..$ in the $\mathrm{CuO}_2$ plane and construction of the bond
lattice out of the centers of the Cu-O-Cu bonds. The solid circles at $\{\mathbf{R}_i\doteq i\}$ (blue) represent the
positions of Cu lattice sites and $\{\mathbf{R}_m\doteq m\equiv ij\}$ (magenta) the positions of bond
centre lattice sites. Alternatively, we denote the bond centre lattice site between $\mathbf{R}_i$ and
$\mathbf{R}_j=\mathbf{R}_i+a\hat{\mu}$  as
$\mathbf{R}_{i\mu}\equiv\mathbf{R}_i+(a/2)\hat{\mu}$ with $\hat{\mu}=+\hat{x},+\hat{y}$. The arrows indicate the 
direction of equivalent
planar spins, with ${\bf S}_m=(\Delta_m \cos\phi_m,\Delta_m\sin \phi_m)$ representing the complex order
parameter $\psi_{ij}\equiv \psi_m=\Delta_m \exp(i\phi_m)$ and antiferromagnetic ordering (shown) of spins
translating into a $d$-wave symmetry gap (long-range order).}
\label{fig.BondLattice}
\end{figure}

\begin{figure}
\begin{center}
\includegraphics[height=6cm]{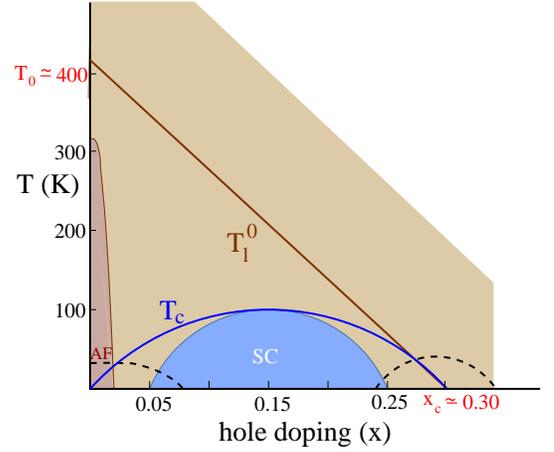}
\end{center}
\caption{ A schematic illustration of the hole doping $x$ and temperature $T$ plane (entire shaded region)
where we assume the functional of Eq.\eqref{Eq.functional} to be applicable.
$T^0_l(x)$ (solid brown line) and $T_c(x)$ (solid blue line) are shown along
with the experimental superconducting (SC) dome and antiferromagnetic (AF) regime at very low hole doping. 
The two arcs shown by dotted lines denote regions where quantum fluctuation effects, as well as other low-energy degrees of 
freedom, such as electronic and spin plus their coupling with pair degrees of freedom, need to be explicitly included in 
the free energy functional. For instance, inclusion of quantum phase fluctuation effects in a minimal level leads to a
$T_c(x)$ curve in agreement with experiment (See Section~\ref{sec.Tc}).}
\label{fig.PhaseDiagram}
\end{figure}

In our phenomenological description, we  hypothesize that a free energy functional similar in structure to that of
Eq.(\ref{Eq.continuumGLfunctional}), but defined on the square planar $\mathrm{CuO}_2$ lattice, describes the properties of
cuprate superconductors for a fairly wide range of hole doping ($x$) and temperature ($T$).
Fig.\ref{fig.BondLattice} shows the square planar lattice schematically, and
Fig.\ref{fig.PhaseDiagram} the region of the $(x,T)$ plane where our phenomenological description is assumed to be applicable. 
The free energy is assumed to
be a functional of the complex spin-singlet pair amplitude
$\psi_{ij} \equiv \psi_m =\Delta_m\exp(i\phi_m)$ where $i$ and $j$ are nearest-neighbor sites of the
square planar Cu lattice and $m$ labels the `bond-center lattice' site located at the center of the bond between
the lattice sites $i$ and $j$ (see Fig.~\ref{fig.BondLattice}). The highly anisotropic cuprate superconductor is
modeled as a weakly coupled stack of  $\mathrm{CuO}_2$ planes in which the superconductivity is believed to
primarily reside and we ignore, as a first approximation, the inter-plane coupling. The free energy functional 
for a single  $\mathrm{CuO}_2$ plane is assumed to have the form
\begin{subequations}\label{Eq.functional}
\begin{eqnarray}
&&\mathcal{F}(\{\Delta_m,\phi_m\})=\mathcal{F}_0(\{\Delta_m\})+\mathcal{F}_1(\{\Delta_m,\phi_m\}),\\
&&\mathcal{F}_0(\{\Delta_m\})=\sum_m \left(A\Delta_m^2 + \frac{B}{2}\Delta_m^4\right),
\label{Eq.functional0}\\
&&\mathcal{F}_1(\{\Delta_m,\phi_m\})=C \sum_{<mn>}  \Delta_m \Delta_n \cos(\phi_m-\phi_n).
\label{Eq.functional1}
\end{eqnarray}
\end{subequations}
A Gor'kov like interpretation of  $\psi_{ij}$ is that it is the average spin-singlet nearest-neighbor Cooper pair
amplitude, i.e. $\psi_{ij}=\langle b_{ij} \rangle/\sqrt{2}=(1/2)\langle a_{i\downarrow}a_{j\uparrow}-
a_{i\uparrow}a_{j\downarrow}\rangle$. The sites $i$
and $j$ are different because strong electron repulsion (symbolized for example by the Mott-Hubbard $U$) disfavors 
on-site  pairing, while the existence of large nearest-neighbor antiferromagnetic spin-spin interaction in the 
parent cuprate is identically equivalent for spin-$\frac{1}{2}$ electrons to attraction between nearest-neighbor 
pairs (i.e. $J_{ij} (\mathbf{S}_i.\mathbf{S}_j-\frac{1}{4}\hat{n}_i\hat{n}_j) = -J_{ij} b^\dagger_{ij} b_{ij}$
with $\mathbf{S}_i$ and $\hat{n}_i$ the electron spin and number operators respectively at the $i$-th site).
This favors the formation of nearest-neighbor spin-singlet pairs. 

The first part $\mathcal{F}_0$  of $\mathcal{F}$ is the sum of identical independent terms each of which is a function of 
only the magnitude $\Delta_m$ of the order parameter on the bond lattice site. Eq.\eqref{Eq.functional0} is a simple form 
for it in the image of Eq.\eqref{Eq.continuumGLfunctional}, with $A$ and $B$ depending, in general, on $x$ and $T$. 
We assume that $B$ is a positive constant independent
of $x$ and $T$ and choose $A(x,T)$ to change sign along a straight 
line $T^0_l(x)$ running from $T= T_0$ at $x=0$ to $T=0$ at $x=x_c$ (see Fig.~\ref{fig.PhaseDiagram}). 
As a first approximation, this line can be 
identified with the pseudogap temperature 
$T^*(x)$ because the magnitude of the local pair ampliitude, $\langle\Delta_m\rangle$, can increase 
dramatically as the temperature crosses
this line, so that $A$ changes from a positive to a negative value. 
The occurrence of superconductivity, characterized by 
a nonzero stiffness for long-wavelength phase fluctuations and the associated superconducting phase coherence,
depends on the phase coupling term, Eq.\eqref{Eq.functional1}. 
If $C$ in Eq.\eqref{Eq.functional1} is taken to be proportional to $x$, the superconducting
transition temperature $T_c$, as calculated in our theory, turns out to be
proportional to $x$ for small values of it, in conformity with what is observed, e.g. the Uemura 
correlation \cite{YJUemura1}. Also, if $C$ is taken to be positive, the transition is to a $d$-wave symmetry superconducting 
state (see Section~\ref{sec.GLFunctional}). We, therefore, make this choice.

We emphasize that the assumed form of the functional and the dependence of the coefficients on $x$ and $T$ are
purely phenomenological, guided by experimental results -- the functional is not derived from a microscopic theory. The
functional satisfies the usual symmetry and stability requirements:  the absence of odd powers of $\psi_m$ ensures
invariance of the free energy under a global change of phase, and the free energy is bounded below for the
chosen positive $B$. Since $\Delta_m = |\psi_m|$ and $\Delta_m\Delta_n \cos(\phi_m-\phi_n) = -(|\psi_m-\psi_n|^2
-\Delta_m^2-\Delta_n^2)/2$, it is
readily seen that the free energy of Eq.(\ref{Eq.functional}) is similar in form to a discretized version of the
GL functional of Eq.(\ref{Eq.continuumGLfunctional}). However, there are important differences between our 
phenomenological approach and the original GL theory -- these differences are discussed in detail in 
Section~\ref{sec.GLFunctional}.

The main objective of our study is to investigate whether the free energy functional defined above provides
a good description of experimental results over a wide range of $x$ and $T$. To this end, we have carried out
several investigations of the thermodynamic behavior of a system whose equilibrium properties are given by 
canonical (thermal) averages with the functional of Eq.(\ref{Eq.functional}) playing the role 
of the `Hamiltonian' or energy function. These calculations have been performed at several levels of sophistication.
We first used simple single-site mean-field theory to obtain qualitative information about the behavior of the system
over a wide range of $x$ and $T$. We also used cluster mean-field theory to obtain more accurate estimates of the
superconducting transition temperature as a function of doping. We used
numerical minimization of the free energy to obtain exact results for 
the properties of the system and the structure of vortices at zero temperature. We also used extensive Monte Carlo (MC)
simulations to obtain exact (modulo finite-size effects) information about the thermodynamic behavior of the system at finite 
temperatures. Since the free energy of Eq.(\ref{Eq.functional}) may be viewed as the Hamiltonian of a two-dimensional
XY model with fluctuations in the magnitudes of the `spins' (see Section~\ref{sec.GLFunctional} for the details of this
analogy), we made use of well-known results about the behavior of the XY model in two dimensions in the analysis
of the data obtained from our MC simulations. Finally, we extended our free energy functional to
include quantum phase fluctuations (see Section~\ref{sec.Tc}) in order to study the effects of these fluctuations on
the transition temperature, and included coupling of the pair degrees of freedom to electrons 
(see Section~\ref{sec.SpectralDensity}) to study the spectral properties of electrons measured in ARPES experiments.
Simple, physically motivated, approximate analytic methods were used in these studies.
The main results obtained from these extensive analytic and numerical calculations are summarized below.

As a starting point, we calculate the superconducting transition temperature $T_c(x)$ and 
the average magnitude of the local pair amplitude, $\langle\Delta_m \rangle$, using single-site mean-field theory for
the model of Eq.\eqref{Eq.functional}. We show that this approximation leads to general 
features of the $x-T$ phase diagram in agreement with experiment. In particular, we find a phase coherent 
superconducting state with $d$-wave symmetry below a parabolic
$T_c(x)$ dome and a phase incoherent state with a perceptible local gap that persists up to a temperature
around $T^0_l(x)$. Further, effects of thermal fluctuations beyond the mean-field level are captured via MC
simulations of the model of Eq.\eqref{Eq.functional} for a finite two-dimensional lattice.
Section \ref{sec.Tc} describes the results for $T_c(x)$ obtained from these simulations. 
The actual values of $A$, $B$ and $C$ used in these calculations are discussed in
Section \ref{sec.GLFunctional}.

The superfluid stiffness $\rho_s(x,T)$ (a quantity measured e.g. via the penetration depth) is calculated in Section
\ref{sec.SuperfluidDensity}. Its doping and temperature dependence compare well with experimental results
\cite{YJUemura2,CBernhard,CNiedermayer,BRBoyce,CPanagopoulos}. The thermally 
averaged local gap $\bar{\Delta}(x,T)\equiv \langle \Delta_m\rangle$ is obtained in Section~\ref{sec.Gap} where we
calculate the temperature $T_\mathrm{ms}(x)$ corresponding to the maximum slope of this quantity with and without the $C$ 
term. This temperature provides a measure of the pseudogap temperature $T^*(x)$.
We use these results to remark on contrasting scenarios \cite{MRNorman1,JLTallon} proposed for the doping dependence of the
pseudogap. We find that there is a contribution to $\bar{\Delta}(x,T)$ that `turns on' at $T_c(x)$, the superconducting
transition temperature. This is obviously connected with persistent observations of two different kinds of
energy gaps in several experiments \cite{MLTacon,JWAlldredge}. We also calculate the ratio 
$2\Delta(x,0)/T_c(x)$ which is observed to be 
generally much larger than the BCS value of about 4 over a wide range of $x$ \cite{OFischer,MKugler}, 
and to vary from system to system
within the cuprate family for the same $x$. Our results rationalize this behavior, which is expected here
since the origins of $\Delta(x,0)$ and $T_c$ are different. 

The contribution of the  pair degrees 
of freedom to thermal properties, such as the specific heat $C_v$, can be obtained from the free energy functional of
Eq.\eqref{Eq.functional}. We briefly report in Section \ref{sec.SpecificHeat} our calculation of
$C_v$ (details are given in a subsequent paper \cite{SBanerjee1}), and find that there are two peaks
\cite{JWLoram1,JWLoram2,JWLoram3,TMatsuzaki} in it, a sharp one 
connected with $T_c$ (ordering of the phase of $\psi_m$) and a relatively broad one (`hump') 
linked to $T^*$ (rapid growth of the  magnitude of $\psi_m$). The former is 
specially sensitive to the presence of a magnetic field, as we find in agreement with 
experiment \cite{AJunod,HWen}. Vortices, which are 
topological singularities in phase, are naturally explored in our approach \cite{MTinkham}. We use the functional
of Eq.\eqref{Eq.functional} to find $\Delta_m$ and $\phi_m$ at different sites $m$ for a $2\pi$ vortex whose core is at 
the center of a square plaquette of Cu lattice sites (Section \ref{sec.Vortex}). 
We find that the vortex changes character from being primarily a phase or Josephson vortex for small $x$ to a more 
BCS-like vortex with a large diminution in the magnitude $\Delta_m$ as one approaches the vortex core for
large $x$. Ref.~\cite{SBanerjee2} describes these results in greater detail.

Experimental information about the pair field $\psi_m$ and its correlations is not obtained directly, but from its
coupling to electrons (e.g. ARPES \cite{ADamascelli,JCCampuzano} and STM \cite{OFischer}), photons (e.g. Raman 
scattering \cite{TPDevereaux} and light absorption \cite{DNBasov}) and neutrons \cite{MAKastner}. We therefore develop 
a theory for the coupling of electrons near the Fermi energy with $\psi_m$ and outline it in Section
\ref{sec.SpectralDensity}. A separate paper \cite{SBanerjee3} describes this approach in detail as well as
the results (e.g. Fermi arcs that are ubiquitous above $T_c$, and the pseudogap for various momentum regions of the Fermi
surface, especially the antinodal region) which compare very well with the results of recent ARPES measurements. 
We present here the results for the antinodal pseudogap filling temperature $T^\mathrm{an}(x)$ and compare it with
the other estimate of the pseudogap temperature, $T_\mathrm{ms}(x)$, obtained in Section~\ref{sec.Gap}. 

The results summarized above establish that the simple free energy functional proposed here provides a 
consistent and qualitatively correct (quantitative in some cases) description of a variery of experimentally
observed properties of cuprate superconductors over a wide range of temperature $T$ and hole concentration $x$.
This is the main conclusion of our study. Section \ref{sec.Discussion} discusses certain generalizations,
applications and limitations of the approach used in our study. Appendices \ref{appendix.MFT},
\ref{appendix.SelfEnergy} and \ref{appendix.XYModel} describe some technical details of the calculations.
 
\section{The Free Energy Functional}\label{sec.GLFunctional}

\subsection{Generalities}\label{Subsec.General}

As noted above, the free energy functional used in our study is phenomenological in nature with experimentally
inspired coefficients. We have deliberately kept it as simple as possible, without violating basic requirements
of symmetry and stability. The form of the functional of Eq.\eqref{Eq.functional} is analogous to that used in 
conventional GL theory. However, our approach is different from the GL
theory in several ways. The form of the free energy functional used in the GL theory of supercoductivity 
and in similar theories of other continuous phase transitions \cite{PMChaikin} can be justified only if the
temperature is close to the transition temperature.
This approach, therefore, is expected to yield quantitatively correct results only
in the vicinity of the superconducting transition. This regime of validity is
ordained by the requirement of smallness and slow spatial variation of the order parameter. 
Our use of the simple, GL-like functional of
Eq.\eqref{Eq.functional} over a broad $(x,T)$ region can not be justified from similar considerations: the validity of
our approach can only be judged a posteriori by comparing its consequences with experiments.
Hence, we have calculated a variety of experimentally measurable 
quantities using the functional of Eq.\eqref{Eq.functional} and 
compared the results with those of experiments. As discussed in detail in subsequent Sections,
we find qualitative (and quantitative in some cases) agreement between the theoretical and experimental results for
a wide variety of properties of cuprate
superconductors. This establishes the usefulness of our phenomenological approach in describing 
the properties of cuprate superconductors over a wide range of $x$ and $T$.

Another important difference between our approach and  conventional GL theory is that the free energy
functional we consider is not coarse-grained in the GL sense. We believe that this is natural because all cuprate
superconductors are characterized by short intrinsic pairing length scales or coarse-graining lengths
($\xi_0\sim 15-20~\AA$ in the cuprates rather than the value of $\sim 10,000~\AA$ for `conventional' 
pure superconductors). We thus use a `nearest-neighbor' coupling of the pair amplitudes defined at the 
sites of the atomic bond lattice in the second term of our functional (Eq.(\ref{Eq.functional1})). Another 
difference between the functional used in our study and that of conventional GL theory is that the 
sign of the coupling constant $C$ in Eq.(\ref{Eq.functional1}) is taken to be {\it positive}, 
so that the pair amplitudes at nearest-neighbor
sites of the bond-center lattice have a phase difference of $\pi$ in the ground state. This difference in sign between
the pair amplitudes on the `horizontal' (in the $x$-direction)  and 'vertical' ($y$-direction) 
bonds of the Cu lattice corresponds to  
the superconducting state having $d$-wave symmetry. This is consistent with the experimental fact that the superconducting 
gap $\Delta_\mathbf{k}$ is  proportional to $(\cos{k_xa}-\cos{k_ya})$, which arises in our description from a combination of 
nearest-neighbor Cooper pairs with relative phases as mentioned above.

Some of the  methods of calculation used in our study are also different from that in the conventional GL theory of
superconductivity in which physical properties are calculated using simple mean field theory. The mean-field results
are expected to be valid if the temperature is outside the so-called 'critical' region~\cite{PMChaikin} around the 
transition temperature where the effects of fluctuations, not included in a mean-field analysis, are important. 
For conventional superconductors with long coherence lengths, the width of the critical region is very 
small, so that mean field theory provides a good description of most of the experimentally observed behavior. This,
however, is not the case for cuprate superconductors with very short coherence lengths and for our model of cuprate
superconductivity. For this reason, we have to
go beyond mean field theory (which provides a qualitatively correct, but not quantitative description of the 
general behavior) and use other methods (such as MC simulations) to obtain accurate results 
for the thermodynamic behavior of our model.

A natural description of the pair amplitude $\psi_m$ is as a planar spin of length 
$\Delta_m$ pointing in a direction that makes an angle $\phi_m$ with a fixed axis.
The thermal (Boltzmann) probability of the length distribution is given 
primarily by $\mathcal{F}_0(\{\Delta_m\})$ of Eq.(\ref{Eq.functional0}) and the term in  
Eq.(\ref{Eq.functional1}) may be thought of as 
the coupling between such `spins'. The temperature $T^*(x)$ can be identified roughly as that at which 
the `spin' at each bond lattice site acquires a sizable length locally without any global ordering of the angles, 
whereas the `antiferromagnetic' ($C>0$) nearest-neighbor interaction leads to global order ($d$-wave
superconductivity) setting in at $T_c$. The two temperatures are well separated for small $x$ because $A$, $B$ and $C$ are so 
chosen that $T^*(x\simeq 0)>> T_c$. The region between $T^*$ and $T_c$ is the pseudogap regime where in the spin 
language, antiferromagnetic short-range correlations grow with decreasing temperature, its length scale diverging at
$T_c$. There is considerable experimental evidence for this view \cite{TTimusk,SHufner,MRNorman1}, though there is also 
the alternative view that $T^*(x)$ is associated with a new long-range order, 
e.g. $d$-density wave (DDW) \cite{SChakravarty} or time reversal symmetry breaking circulating currents \cite{CMVarma2}.

The BCS theory and conventional GL theory in which the `spin' formation and ordering 
temperatures are the same are limiting cases of this scenario. Something like this is expected to happen in cuprates near 
$x_c$ (Fig.\ref{fig.PhaseDiagram}) as also follows from our functional. The state below $T_c$ has nonzero order
parameter $\langle\psi_m\rangle$ for a system above two dimensions, and is a Berezinskii-Kosterlitz-Thouless (BKT)
\cite{VLBerezinskii,JMKosterlitz1,JMKosterlitz2} bound 
vortex state with quasi long range order in two dimensions, in which case $T_c$ is identified with the vortex unbinding 
temperature $T_\mathrm{BKT}$. In the former case, the order parameter is the sublattice magnetization
$\Delta_d(x,T)=|\langle \psi_m\rangle|$ with a $\mathbf{k}$-dependent gap 
$\Delta_\mathbf{k}=  (\Delta_d/2)(\cos{k_xa}-\cos{k_ya})$. The interlayer 
coupling can be described, a la Lawrence and Doniach \cite{WELawrence}, by adding say a nearest-neighbor coupling between
`spins' on different layers to our functional in Eq.\eqref{Eq.functional}. Since this is in practice relatively small 
(the measured anisotropy ratio in Bi2212 is about 100, for example \cite{TSchneider}), it makes very little difference 
quantitatively to most of our estimates which generally neglect this coupling. For instance $T_c$ calculated by
estimating the BKT transition temperature ($T_\mathrm{BKT}$) from MC simulation of the two-dimensional model of
Eq.\eqref{Eq.functional} (See Section \ref{sec.Tc}) is expected to be very close to
the actual transition temperature in the anisotropic 3D model with such small interlayer coupling.

A conventional GL theory of cuprate superconductivity would involve a functional 
similar to  that in Eq.(\ref{Eq.continuumGLfunctional})  
(but with additional terms allowed by symmetry) with $\psi({\bf r})$ the  $d$-wave superconducting order parameter,
and the coefficient so chosen that a mean-field treatment of the free energy leads to a dome-shaped $T_c(x)$ curve
similar to that found in experiments. However,
a mean-field treatment and the conclusions obtained from it would not be reliable because  
of the smallness of the superconducting 
coherence length in the cuprates and consequent large fluctuation effects.  In particular, the pseudogap temperature 
$T^*$ (which is much  larger than $T_c$ for small $x$, and goes to $T_c$ as  $x$ increases) would  be 
absent in such a theory. By contrast, we assume here that the basic low-energy Cooper  pair degree of freedom in the 
cuprates is the bond pair, give a physical meaning to $T^*$ as  a pair magnitude crossover temperature, and describe 
the regime between $T^*$ and $T_c$  as  one in which the correlation length associated with  superconducting fluctuations of 
$d$-wave symmetry grows and diverges at $T_c$. The effect of these fluctuations is found to  be crucial for many physical 
properties, e.g. the Fermi arc phenomenon, and the filling of  the antinodal pseudogap as $T$ rises to $T^*$. The 
superconducting order with $d$-wave  symmetry that sets in at $T_c$ is an emergent collective effect, arising from the 
short-range  $\psi^*_m \psi_n$ interaction, much as long-range Neel order arises from an  antiferromagentic coupling 
between nearest-neighbor spins.

GL theories for cuprates have been proposed by a large number of authors, arising either out of a 
particular model for electronic behavior and often coupled with the assumption of a particular
`glue' for binding electrons into pairs \cite{GBaskaran,MDrzazga,LTewordt}, or out of lattice symmetry 
considerations \cite{DLFeder,AJBerlinsky}. The functional in Eq.\eqref{Eq.functional} is consistent with square lattice 
symmetry and, in principle, does not assume any particular electronic approach (weak coupling 
or strong correlation, for example) or a mechanism for the `glue'. However, some of the 
properties of the coefficients are natural in a strong electron correlation framework. For example, mobile holes in such 
a system can cause a transition between a state in which there is a Cooper pair in the $x$ directed $ij$ bond 
(Fig.\ref{fig.BondLattice}) to one in which the Cooper pair is in an otherwise identical but $y$ directed 
bond $jk$ nearest to it (or vice versa), thus leading to a nonzero term $\mathcal{F}_1$ in
Eq.\eqref{Eq.functional}. This is probably connected with the observed~\cite{EPavarini} empirical correlation between
$T_c$ and the diagonal or next-nearest-neighbor hopping amplitude of electrons in the Cu lattice. 
 
\subsection{Parameters of the Functional} \label{subsec.GLParameters}

 The coefficients $A$, $B$ and $C$ are chosen to be consistent with experiments. Specifically, the coefficients are as 
follows:

\begin{subequations}\label{Eq.GLparameters}
\begin{eqnarray}
A(x,T)&=& A_0 \left[T-T_0 \left(1-\frac{x}{x_c}\right)\right] e^{T/T_p},\label{Eq.A}\\
B&=&B_0T_0, \label{Eq.B}\\
C(x)&=&xC_0T_0, \label{Eq.C}
\end{eqnarray} 
\end{subequations}
We require $\Delta_m$ to have dimensions of
energy $[E]$ (or temperature for Boltzmann constant $k_B=1$) and hence $A_0$, $B_0$ and $C_0$ have dimensions
of $[E]^{-2}$, $[E]^{-4}$ and $[E]^{-2}$ respectively. They are rewritten in terms of $T_0$ as well as three
dimensionless parameters $f$, $b$ and $c$ so that $\mathcal{F}$ carries dimension of energy as well. We thus
have, $A_0=(f/T_0)^2$, $B_0=b(f/T_0)^4$ and $C_0=c(f/T_0)^2$. We choose $b$ and $c$ to have values of order unity and fix 
them for different hole doped cuprates by comparing $\Delta_0(x)$, $T^*(x)$ and $T_{c}^\mathrm{opt}$ obtained from the theory with experiments (see below for details).

The two temperature dependent parts of $A$ as given above arise as follows. The part $[T-T_0(1-x/x_c)]$
reflects our identification of the zero of $A(x,T)$ with the pseudogap temperature and the experimental
observation that the pseudogap region extends downwards nearly linearly from $T = T_0$ at $x=0$ to  $T=0$ for
$x=x_c$. The relation between this straight line $T^0_l(x)$, the experimental $T^*(x)$ and the related quantities
$T^{0,1}_\mathrm{ms}(x)$ (obtained  from a maximum slope criterion, Section \ref{sec.Gap}) as well as $T^\mathrm{an}(x)$ 
(obtained from the antinodal gap filling criterion for the electron spectral function, Section
\ref{sec.SpectralDensity}) is shown in Fig.\ref{fig.PseudogapT} and
Fig.\ref{fig.AntinodalGap}. The exponential factor $e^{T/T_p}$ suppresses $\bar{\Delta}(x,T)$ at high
temperatures ($T>>T^0_l(x)$) with respect to its  temperature independent equipartition value
$\sqrt{T/A(x,T)}$ which will result from the classical functional (Eq.\eqref{Eq.functional}) being used
well beyond the near proximity of any critical temperature where it is valid. Such a suppression is natural in
a degenerate Fermi system; the relevant local electron pair susceptibility is rather small above the pair
binding temperature and below the degeneracy temperature. The temperature scale $T_p$ is of order $T_0$, this
being the energy scale for pair binding. We take it to be $T_0$ unless stated otherwise. In all the
calculations below, we choose $x_c=0.3$ and $b=0.1$ (except in Fig.\ref{fig.TcQuantum}(b)). $b$ along with $T_p$ 
controls the temperature dependence
of $\bar{\Delta}(x,T)$, especially the decrease of $\bar{\Delta}(x,T)$ across the `pseudogap temperature' line
$T^*(x)$ and other details such as the height of the specific heat hump around $T^*(x)$. Values of $f$, $c$ and
$T_0$ can be fixed for a variety of cuprates by comparing zero temperature gap $\Delta_0(x)$, $T^*(x)$ and
$T_c^\mathrm{opt}$ with experiments. For example, a choice of parameters, roughly suitable for Bi2212, which has an 
experimental $T_c^\mathrm{opt}\simeq 91$ K, gives $f\simeq 1.33$, $c\simeq0.3$ with $\Delta_0(x=0)\simeq 82$ meV,
$T_0\simeq400$ K and $T_\mathrm{BKT}^\mathrm{opt}\simeq 72$ K ($T_c^\mathrm{opt}\simeq 110$ K from single site mean
field theory, see Section \ref{sec.Tc}). Unless otherwise stated, we have used the above choice of
parameter values in the rest of the paper.
         
\section{Superconducting Transition Temperature $T_c(x)$}\label{sec.Tc}

The superconducting state is characterized by macroscopic phase coherence. For superconductivity in cuprates
described by the functional (Eq.\eqref{Eq.functional}) this means a nonzero value for the superfluid
stiffness or superfluid density $\rho_s(x,T)$ given by the formula \cite{WYShih}, 
\begin{eqnarray}
&&\rho_s=-\frac{C}{2N_b}\langle \sum_{m,\mu}\Delta_m\Delta_{m+\mu}\cos(\phi_m-\phi_{m+\mu})\rangle\nonumber \\
&&-\frac{C^2}{2N_bT}\sum_\mu\langle (\sum_m \Delta_m\Delta_{m+\mu}\sin(\phi_m-\phi_{m+\mu}))^2\rangle \nonumber \\
\label{Eq.sfldensity}
\end{eqnarray}
where the subscript $m+\mu$ refers to $\mathbf{R}_m+l\hat{\mu}$ with $\hat{\mu}$ running over $x$ and $y$ directions
in the bond lattice coordinate system (rotated by $45^0$ with respect to the $x$-axis shown in Fig.\ref{fig.BondLattice}),
$l=a/\sqrt{2}$ is the spacing of the bond lattice, and $N_b$ is number of sites in the bond lattice
($N_b=2N$). The superconducting transition temperature $T_c(x)$ is the highest 
temperature at which $\rho_s(x,T)$ is nonzero. 
We use this fact to obtain $T_c(x)$ in 
single-site and cluster mean-field theories (the relevant details are summarized in Appendix
\ref{appendix.MFT}). As  mean field approximations are known~\cite{PMChaikin} to overestimate the transition temperature,
we treat the effect of fluctuations in the model of (Eq.\eqref{Eq.functional}) through MC 
simulations. In these simulations, the standard Metropolis sampling
scheme \cite{MEJNewman} has been used for planar spins $\{\mathbf{S}_m=(\Delta_m\cos\phi_m,\Delta_m\sin\phi_m)\}$, whose
lengths are controlled mainly by $\mathcal{F}_0$ (Eq.\eqref{Eq.functional0}). Simulations have been carried out for a
$100\times100$ square lattice (bond lattice) with periodic boundary condition. Typically, $10^5$ MC steps per spin have been
used for equilibration and measurements were done for next $3\times10^5$ ($6\times10^5$ in some cases) MC
steps per spin. Simulations were done for the doping range $0-0.4$ at various temperatures.

\begin{figure}
\begin{center}
\begin{tabular}{c}
\includegraphics[height=6cm]{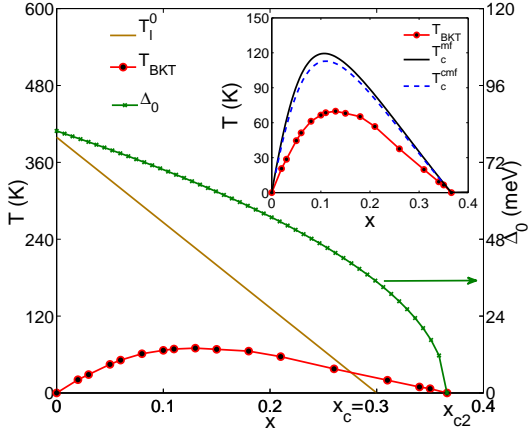}
\end{tabular} 
\end{center}
\caption{ Doping dependence of different temperature scales ($T^0_l$ and
$T_\mathrm{BKT}$) and the zero temperature gap $\Delta_0$ (Eq.\eqref{Eq.gap0}) are shown in the main plot. 
Inset: Comparison of the $T_c$'s obtained 
from single-site mean-field theory and cluster mean-field theory ($T_c^\mathrm{mf}$ and $T_c^\mathrm{cmf}$ respectively)
(see Appendix \ref{appendix.MFT}) with the BKT transition temperature $T_\mathrm{BKT}$ obtained from MC
simulation, as discussed in the text.}
\label{fig.Tc}
\end{figure}

In our two-dimensional model, true long-range order is destroyed by thermal 
fluctuations, but there is nonzero superfluid stiffness due to 
vortex-antivortex binding (the BKT transition 
\cite{VLBerezinskii,JMKosterlitz1,JMKosterlitz2}) below a temperature $T_\mathrm{BKT}$. We calculate the superfluid 
stiffness in the MC simulation using the formula of
Eq.\eqref{Eq.sfldensity} and use it in conjunction with the Nelson-Kosterlitz criterion \cite{DRNelson}
\begin{eqnarray}
\frac{\rho_s(T_\mathrm{BKT})}{T_\mathrm{BKT}}=\frac{2}{\pi} \label{Eq.NelsonJump}
\end{eqnarray}
based on the BKT theory to obtain the vortex binding temperature $T_\mathrm{BKT}(x)$, which is identical to
$T_c(x)$ in 2D. The above criterion, appropriate for a 
fixed length XY  model or equivalently a low fugacity 2D vortex gas, might not give an accurate estimate of 
$T_\mathrm{BKT}$ for the model of Eq.\eqref{Eq.functional} in the extreme
overdoped regime close to $x=x_c$ due to large fluctuations in the magnitudes $\Delta_m$ \cite{DBormann}. $T_\mathrm{BKT}$ 
obtained using Eq.\eqref{Eq.NelsonJump} should presumably be quite accurate in the underdoped and optimally doped regions 
where the magnitudes effectively become `frozen' since $T^*(x)>>T_c(x)$ resulting in a description of the model 
(Eq.\eqref{Eq.functional}) in terms of an effective fixed-length XY model (Appendix \ref{appendix.XYModel}) close to
the superconducting transition. These results are shown in Fig.\ref{fig.Tc}. Results for the temperature dependence of the 
superfluid stiffness are presented in Section
\ref{sec.SuperfluidDensity}.

\begin{figure}
\begin{center}
\begin{tabular}{c}
\includegraphics[height=6cm]{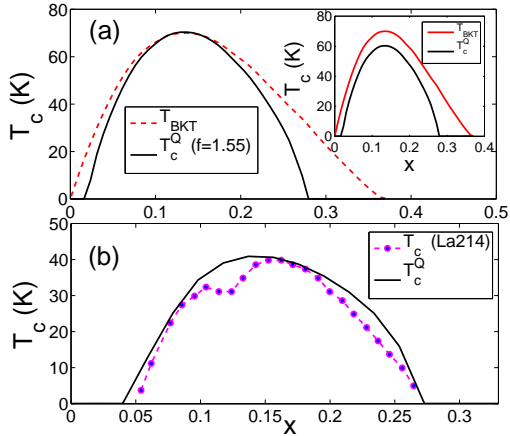}
\end{tabular} 
\end{center}
\caption{{\bf (a)} Effect of quantum fluctuation on $T_c(x)$ curve of Fig.\ref{fig.Tc} for $V_0=0.09T_0$. The
quantum fluctuation renormalizes $T_c$ to $T_c^Q$ throughout the whole $x$ range (Inset). In the main figure, we have
taken $f=1.55$ to change the temperature scale $T_0$ ($=460$ K) while keeping $\Delta_0(x=0)=82$ meV
(Section \ref{subsec.GLParameters}) so that the optimal value of $T_c^Q$ matches that of $T_\mathrm{BKT}$ in 
Fig.\ref{fig.Tc}. 
{\bf (b)} A reasonably good comparison can be obtained with experimental $T_c(x)$ curve 
for La214 with
following choice of parameters (Section \ref{subsec.GLParameters}): $x_c=0.345$, $c=0.33$, $b=0.155$,
$f=1.063$, $T_p=T_0$ and $V_0=0.15T_0$ with $\Delta_0(x=0)=82$ meV. This choice implies $T_0=400$ K. The dip
of the experimental $T_c$ around $x\sim0.12$ is due to the $1/8$ `stripe anomaly' \cite{ARMoodenbaugh} which is out of
the scope of the present functional of Eq.\eqref{Eq.functional} (see discussion in Section \ref{sec.Discussion}).}
\label{fig.TcQuantum}
\end{figure}

The calculated $T_c$ curve is approximately of the same parabolic shape as that found experimentally. 
The causes for the 
qualitative disagreement at both ends (see Fig.\ref{fig.PhaseDiagram}) are not difficult to understand.
For very small $x$, as well as for $x$ near $x_c$, our free energy functional needs to be extended by including
quantum phase fluctuation effects. For such values of $x$, zero point fluctuations are important because the phase
stiffness is small. Additionally, low-energy mobile electron degrees of freedom need to be considered
explicitly for $x$ near $x_c$.
To include quantum phase 
fluctuation effects, we supplement the GL functional of Eq.\eqref{Eq.functional} with the following term that describes 
quantum fluctuations of phases ($\phi_m$) at a minimal level \cite{SDoniach,ERoddick,MFranz}: 
\begin{eqnarray}
\mathcal{F}_Q(\{\hat{q}_m\})&=&\frac{1}{2}\sum_{mn} \hat{q}_m V_{mn} \hat{q}_n \label{Eq.functionalQ}
\end{eqnarray}
Here $\hat{q}_m$ is the Cooper pair number operator at site $m$, and $\phi_m$ in Eq.\eqref{Eq.functional1}
should be treated as a quantum mechanical operator $\hat{\phi}_m$, canonically conjugate to $\hat{q}_m$ so
that $[\hat{q}_m,\hat{\phi}_n]=i\delta_{mn}$ \cite{RFazio}. We take the simplest possible form for $V_{mn}$ i.e. 
$V_{mn}=V_0\delta_{mn}$ for the purpose of demonstrating the effect of quantum fluctuations on the $T_c(x)$ curve
(Fig.\ref{fig.TcQuantum}), where $V_0$ is the strength of on-site Cooper pair interaction. We have obtained a single-site 
mean field estimate of $T_c(x)$, namely $T_c^Q(x)$, including the effect of $\mathcal{F}_Q$ as shown in
Fig.\ref{fig.TcQuantum} and discussed in Appendix \ref{appendix.MFT}. As it is well known, mean field theory
overestimates the value of the transition temperature. Hence to compare $T_c^Q(x)$ with $T_\mathrm{BKT}(x)$
of Fig.\ref{fig.Tc} as well as with the experimental $T_c(x)$ curve, we scale the $T_c^Q$ calculated using 
Eq.\eqref{Eq.MFTTcQuantum} by a factor $\sim 0.6$ in Fig.\ref{fig.TcQuantum}. This factor has been estimated by calculating 
the ratio $T_\mathrm{BKT}(x)/T_c^\mathrm{mf}(x)$ from Fig.\ref{fig.Tc} (inset). Quantitative agreement for $T_c $ for a 
specific cuprate, $\mathrm{La_{2-x}Sr_xCuO_4}$ is possible with a particular choice of parameters as shown in 
Fig.\ref{fig.TcQuantum}(b). In this extension of the model, we have ignored the long-range nature of the Coulomb 
(or charge) interactions, as well as Ohmic dissipation. It has been argued \cite{ERoddick} that these two factors together 
result in a fluctuation spectrum similar to the one obtained in an approximation that ignores both, but retains the 
short-range part of the charge interaction. 

In the remaining parts of the paper, we do not consider quantum phase
fluctuations since they modify the results qualitatively only in the extremely underdoped and overdoped regions
by aborting the superconducting transition as the phase stiffness $\rho_s(0)$ becomes small (see
Fig.\ref{fig.SuperfluidDensity}) at these two extremes in 
our model. In the rest of the $x$ range, these effects are expected to renormalize \cite{JVJose} the values of the 
parameters of the functional of Eq.\eqref{Eq.GLparameters}. We assume that such renormalizations are implicit in our choice 
of the parameters $A$, $B$ and $C$ in tune with experimental facts (see Section \ref{subsec.GLParameters}).       

\section{Superfluid Density $\rho_s(x,T)$} \label{sec.SuperfluidDensity}
   
\begin{figure}
\begin{center}
\begin{tabular}{c}
\includegraphics[height=6cm]{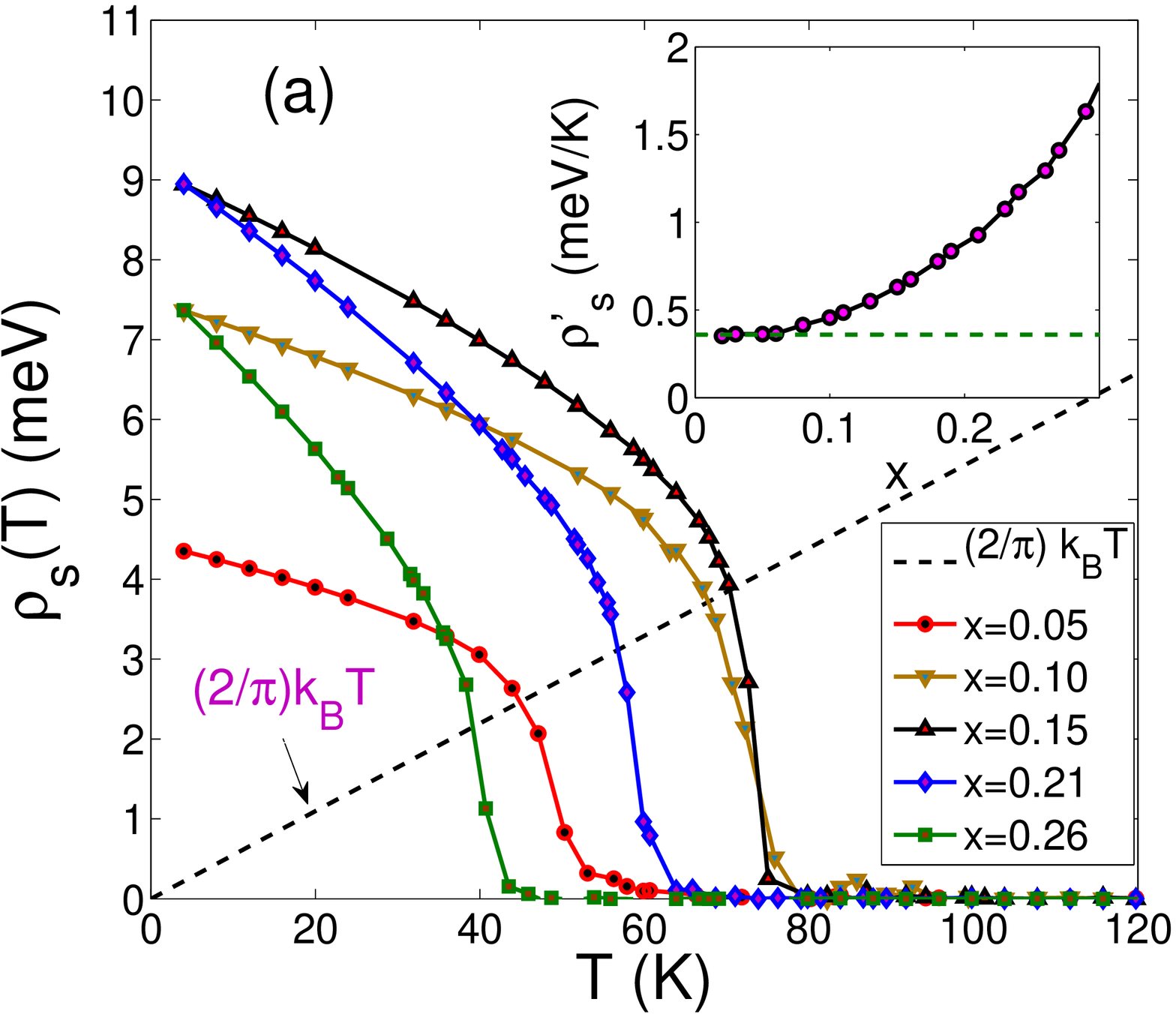}\\
\includegraphics[height=5.5cm]{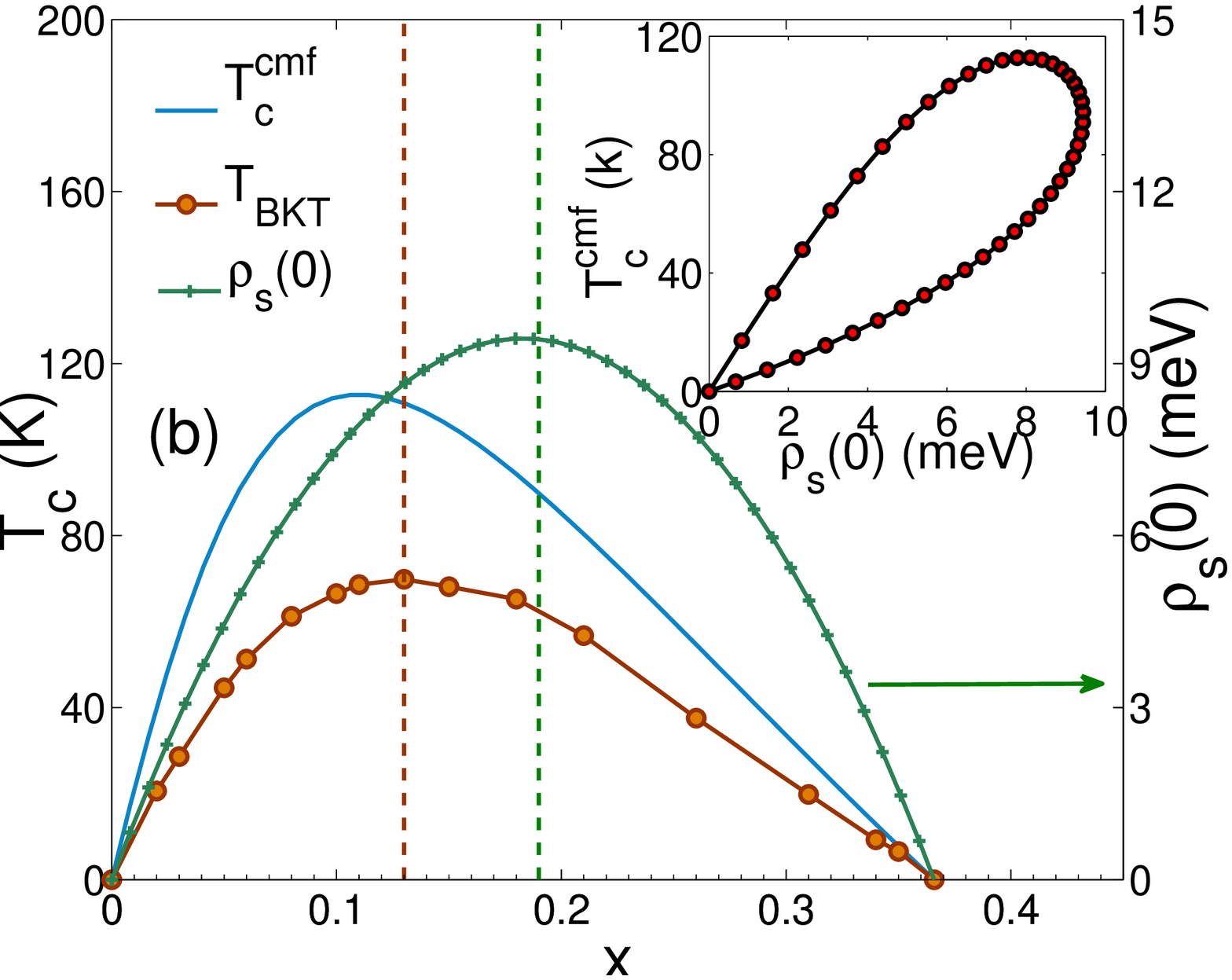}
\end{tabular}
\end{center}
\caption{{\bf (a)} Calculated finite temperature superfluid density for different $x$ values. The dashed line
corresponds to the size of universal Nelson-Kosterlitz jump (Eq.\eqref{Eq.NelsonJump}) expected at a BKT
transition. $T_\mathrm{BKT}(x)$ has been obtained from the intersection of this line with $\rho_s(x,T)$ vs.
$T$ curves. Inset: $\rho_s^\prime(x)$, estimated by fitting $\rho_s(x,T)$ vs. $T$ with a linear form, 
$\rho_s(x,T)=\rho_s(x,0)-\rho_s^\prime(x) T$. 
{\bf (b)} Zero temperature superfluid density $\rho_s(x,0)$, as a function of $x$, 
compared with $T_\mathrm{BKT}(x)$ and $T_c^\mathrm{cmf}$(x). The superfluid density has been expressed in units of
energy (meV) as appropriate in 2D. Vertical dashed lines indicate $x$'s corresponding to optimal values of
$\rho_s(x,0)$ and $T_\mathrm{BKT}(x)$. The inset shows the `Uemura plot' \cite{YJUemura1,YJUemura2,CNiedermayer}, 
$T_c(x)$ vs. $\rho_s(x,0)$. The initial part of the upper branch corresponds the underdoped region,
where the Uemura relation was inferred \cite{YJUemura1} originally. The subsequent decrease of $\rho_s(x,0)$ along
with $T_c$ in the overdoped regime (lower branch) is observed for example in $\mathrm{Tl_2Ba_2CuO_{6+\delta}}$
\cite{YJUemura2,CNiedermayer}.} 
\label{fig.SuperfluidDensity}
\end{figure}

As mentioned above, we have evaluated the superfluid density $\rho_s$ at finite temperatures using 
Eq.\eqref{Eq.sfldensity} by MC simulation of our model (Eq.\eqref{Eq.functional}). The results are
discussed below along with mean-field results. As we have mentioned in Section \ref{sec.Tc}, the transition temperature
$T_\mathrm{BKT}$ can be estimated from the universal Nelson-Kosterlitz
jump of Eq.\eqref{Eq.NelsonJump}, where $\rho_s(T)=0$ above $T_c$.
We show the results for finite temperature superfluid density in Fig.\ref{fig.SuperfluidDensity}(a).

The zero temperature superfluid density can be calculated easily from the ground state energy change due to a
phase twist (a `spin wave') and is given by 
\begin{equation}
\rho_s(x,0)=C\Delta_0^2(x)
\end{equation}
where $\Delta_0^2(x)$ is obtained from Eq.\eqref{Eq.gap0} (see Section \ref{sec.Gap}). Evidently,
$\rho_s(x,0)\propto x$ for small $x$ (as is implicit in the choice of $C$). $T_c(x)$, of course, is also proportional to 
$x$ for small $x$, as can
be easily verified from Eq.\eqref{Eq.Tc0} (see Appendix A), which gives a quite accurate estimate of $T_c$ for
low hole doping. Hence, the Uemura relation \cite{YJUemura1} is seen explicitly to be satisfied for this
choice of $C$. In
Fig.\ref{fig.SuperfluidDensity}(b) we plot $\rho_s(x,0)$ as a function of $x$ along with $T_c(x)$. $\rho_s(x,0)$ 
initially increases with $x$ to reach a maximum value (so that $\frac{d\rho_s(0)}{dx}=0$) slightly  on the
overdoped side at $x=x_{c_2}/2$ and then ultimately drops to zero at $x_{c_2}$ as $T_c$ also does (see
Fig.\ref{fig.Tc}), but the optimal $T_c(x)$ and optimal $\rho_s(x,0)$ appear, in general, at two different
values of doping ($x_{c_2}/2>x_\mathrm{opt}$ for the present choice of parameters). A similar behavior is
observed in experimental studies of muon-spin depolarization rate,  $\sigma_0\propto\rho_s(x,0)$ of some
cuprates which can be sufficiently overdoped \cite{YJUemura2,CBernhard}. The  depolarization rate depends on
the local magnetic field at the location of the muon; this has been shown to be proportional to the superfluid
stiffness which controls the magnetic response of the superfluid \cite{JESonier}. We also plot $T_c(x)$ as a function of
$\rho_s(x,0)$ (`Uemura plot', inset of Fig.\ref{fig.SuperfluidDensity}(b))which compares well with
experimental plots of $T_c$ vs. $\sigma_0$, measured at low temperatures and shown in
Refs.\onlinecite{YJUemura2,CNiedermayer}.

At low temperatures the calculated $\rho_s(x,T)$ decreases linearly with $T$ from its zero temperature value
i.e. $\rho_s(x,T)=\rho_s(x,0)-\rho_s^\prime(x)~T$; the coefficient of the linear term, namely
$\rho_s^\prime(x)$ remains more or less independent of $x$ for small $x$ and approaches a constant value as
$x\rightarrow 0$ on the underdoped side. The same trend can be observed in the experimental data
\cite{BRBoyce,CPanagopoulos} for in-plane magnetic penetration depth $\lambda_{ab}$, where
$\lambda_{ab}^{-2}\propto \rho_s$. It is interesting that a model for superconductivity such as ours, which does not
explicitly include electron degrees of freedom leads to a linear decrease \cite{TOhta,EWCarlson}, in the light of the fact 
that the linear dependence has been attributed to thermal, nodal quasiparticles of the $d$-wave superconductor
\cite{PALee}.   
 
\section{Average Local Gap $\bar{\Delta}(x,T)$ and the Pseudogap} \label{sec.Gap}

 The energy gap $\Delta_m$ is a thermodynamic variable with a certain probability distribution given by the 
functional of Eq.\eqref{Eq.functional}. There is no direct measurement of the energy gap, unlike that of $T_c$
or of the superfluid stiffness discussed in Sections \ref{sec.Tc} and \ref{sec.SuperfluidDensity}. The
information about the energy gap is obtained via the coupling of the gap (or more precisely, of electron pairs
giving rise to the gap) to electrons, photons, neutrons etc. 
In this section, we compute the thermodynamically averaged local gap $\bar{\Delta}(x,T)=<\Delta_m>$ and compare 
our results with the broadly observed trends for gaps as inferred from a number of measurements on a variety of cuprates. 
These trends are for the pseudogap as a function of hole doping $x$, and for the ratio of the zero temperature gap to the 
pseudogap temperature $T^*(x)$ as well as to the directly measured superconducting $T_c$.  
     
Fig.\ref{fig.Gap} shows the dependence of $\bar{\Delta}(x,T)$, calculated in single site mean field
theory (see Appendix \ref{appendix.MFT}), on temperature for different values of the hole doping $x$. We
have checked that the values of $\langle \Delta_m \rangle$ obtained
from MC simulations are quite similar to the mean-field results, the main difference being that the
singularity of the mean-field values at $T_c(x)$ is smoothed out in the MC results. Note that the quantity
$\Delta_m=|\psi_m|$ is {\it not} the order parameter for superconductivity and its average $\bar{\Delta}(x,T)$ can be 
(and is) nonzero at temperatures above $T_c$. The average gap
increases smoothly as $T$ decreases; the increase can be rather abrupt or gradual, depending on the parameters
(see Fig.\ref{fig.Gap}(b)). The part in $\bar{\Delta}(x,T)$ `turning on' at $T_c$ is generally small. The zero temperature 
gap $\Delta_0(x)\equiv \bar{\Delta}(x,0)$, is the sum of these two, a gap which would have been there even in the absence 
of phase coherence (shown by the dotted line and calculated from $\tilde{\Delta}=\langle \Delta_m\rangle_0$,
where the thermal average is evaluated using the single site term $\mathcal{F}_0$ of Eq.\eqref{Eq.functional})
and another, due entirely to phase coherence. 

\begin{figure}
\begin{center}
\begin{tabular}{c}
\includegraphics[height=6cm]{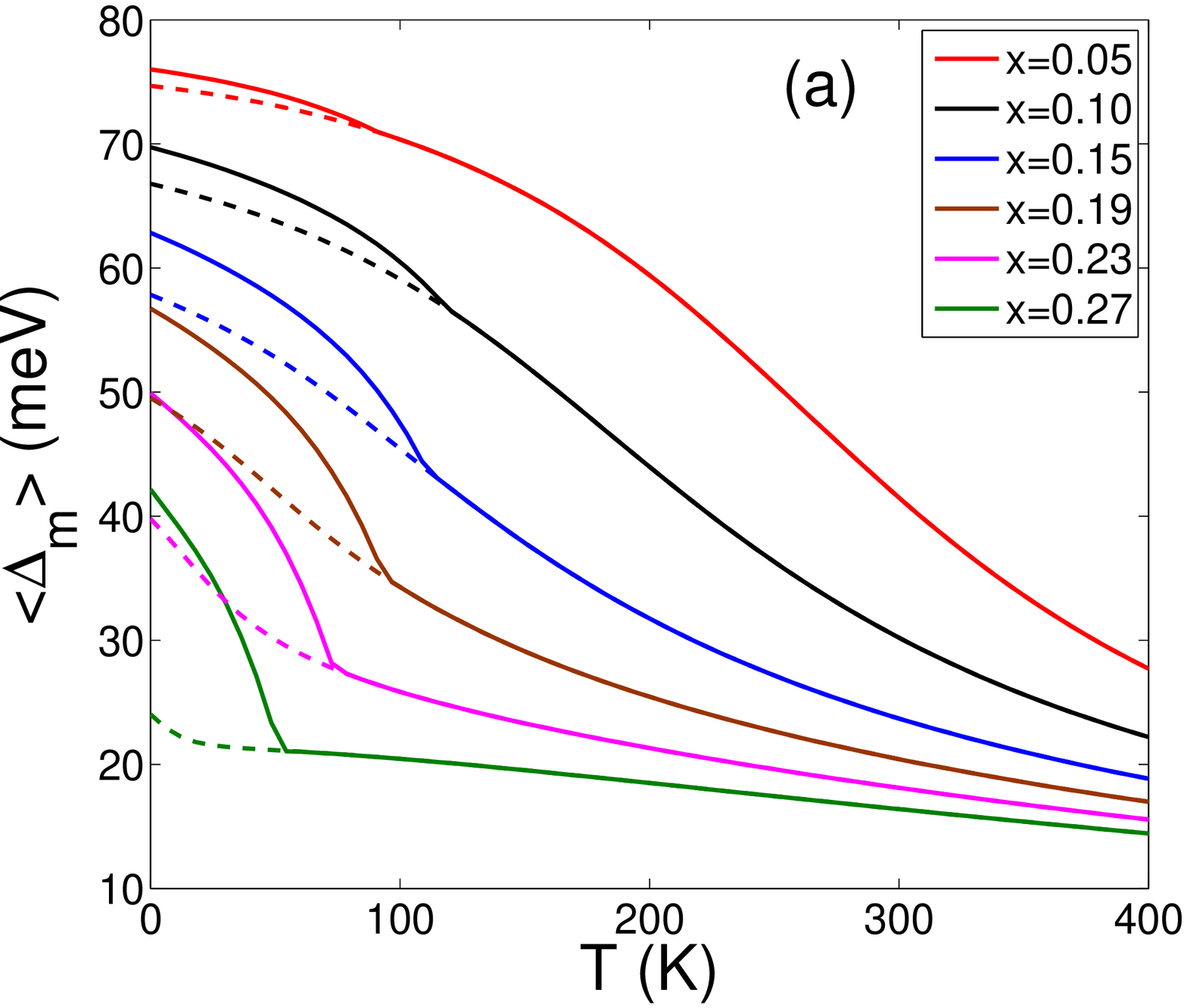}\\ 
\includegraphics[height=6cm]{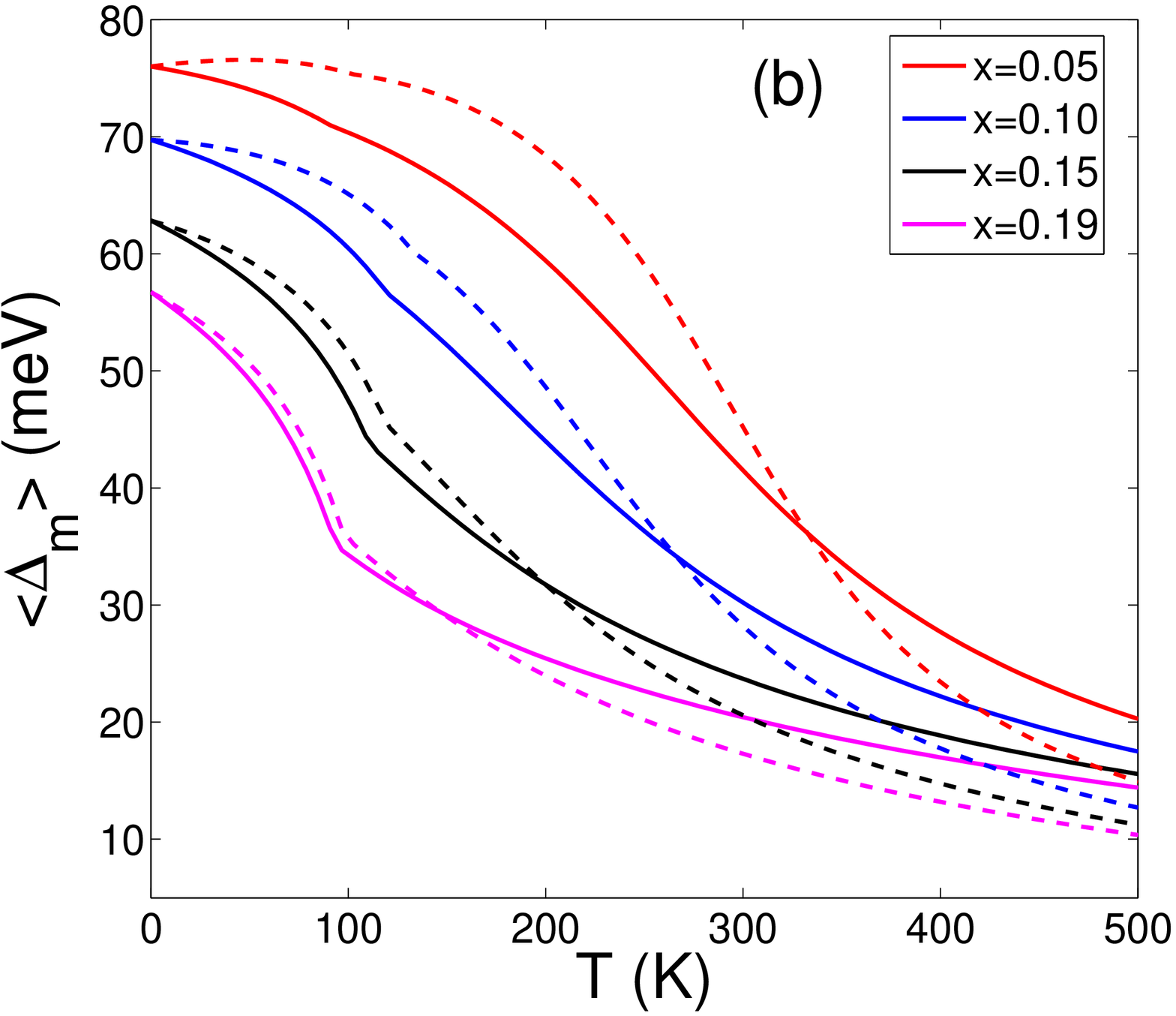}
\end{tabular}
\end{center}
\caption{Panel {\bf (a)} shows the onset of second gap feature in $\bar{\Delta}=\langle \Delta_m\rangle$ at $T_c$ due to the presence 
of the $C$ term in Eq.\eqref{Eq.functional}. The dashed lines compares $\tilde{\Delta}=\langle\Delta_m\rangle_0$
with $\bar{\Delta}$(see text). Panel {\bf (b)} compares the temperature dependence of $\bar{\Delta}$ for $T_p=T_0$ 
(solid lines) and for $T_p=0.65T_0$ (dashed lines). $\bar{\Delta}$ changes much
more rapidly, especially in the underdoped side, with decreasing temperature across $T^{0,1}_\mathrm{ms}(x)$ for
the second case. The results shown here and in Fig.\ref{fig.PseudogapT} were
obtained from single-site mean-field theory.}
\label{fig.Gap}
\end{figure}

\begin{figure}
\begin{center}
\begin{tabular}{c}
\includegraphics[height=7.5cm,angle=-90]{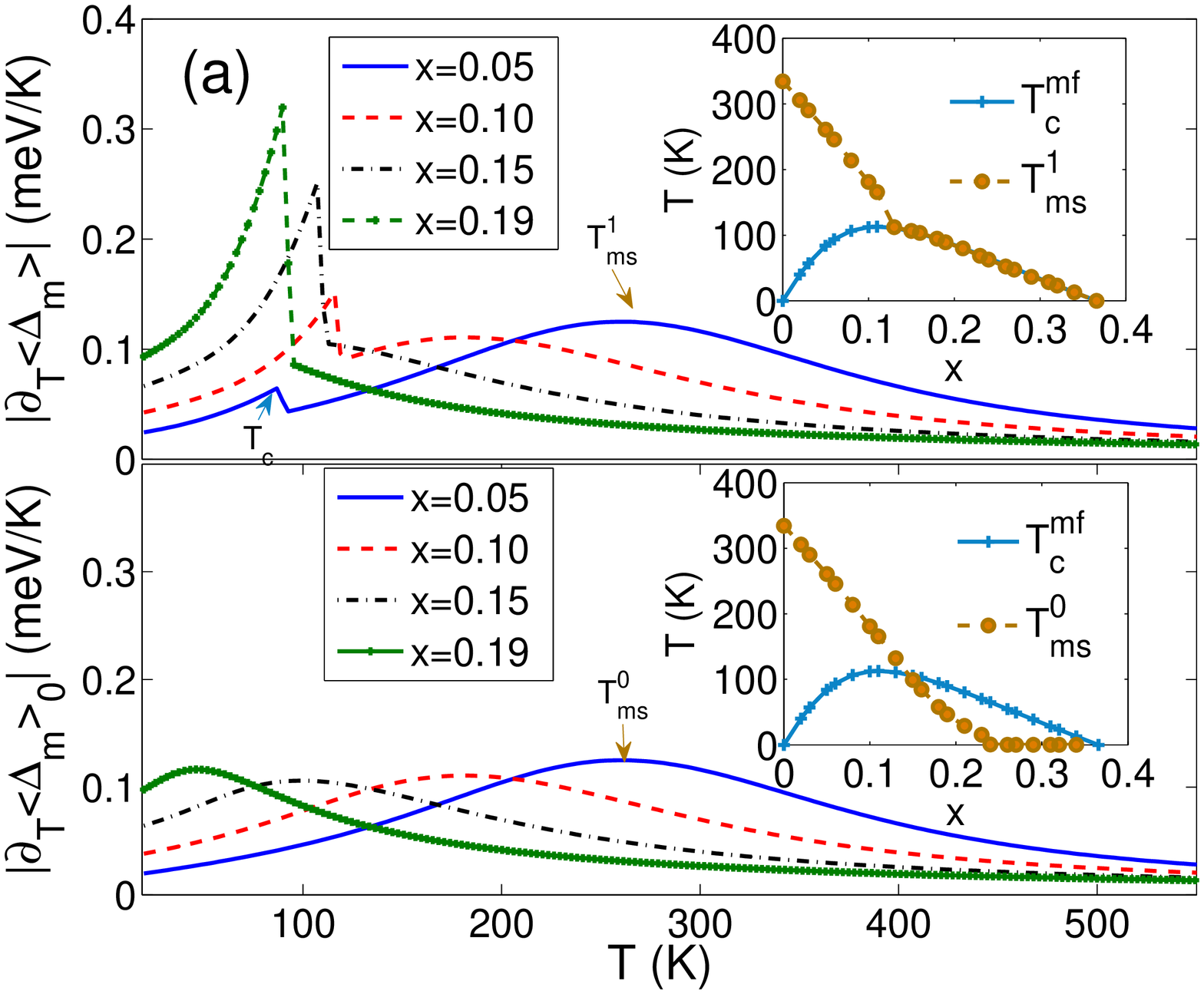}\\ 
\includegraphics[height=7.5cm,angle=-90]{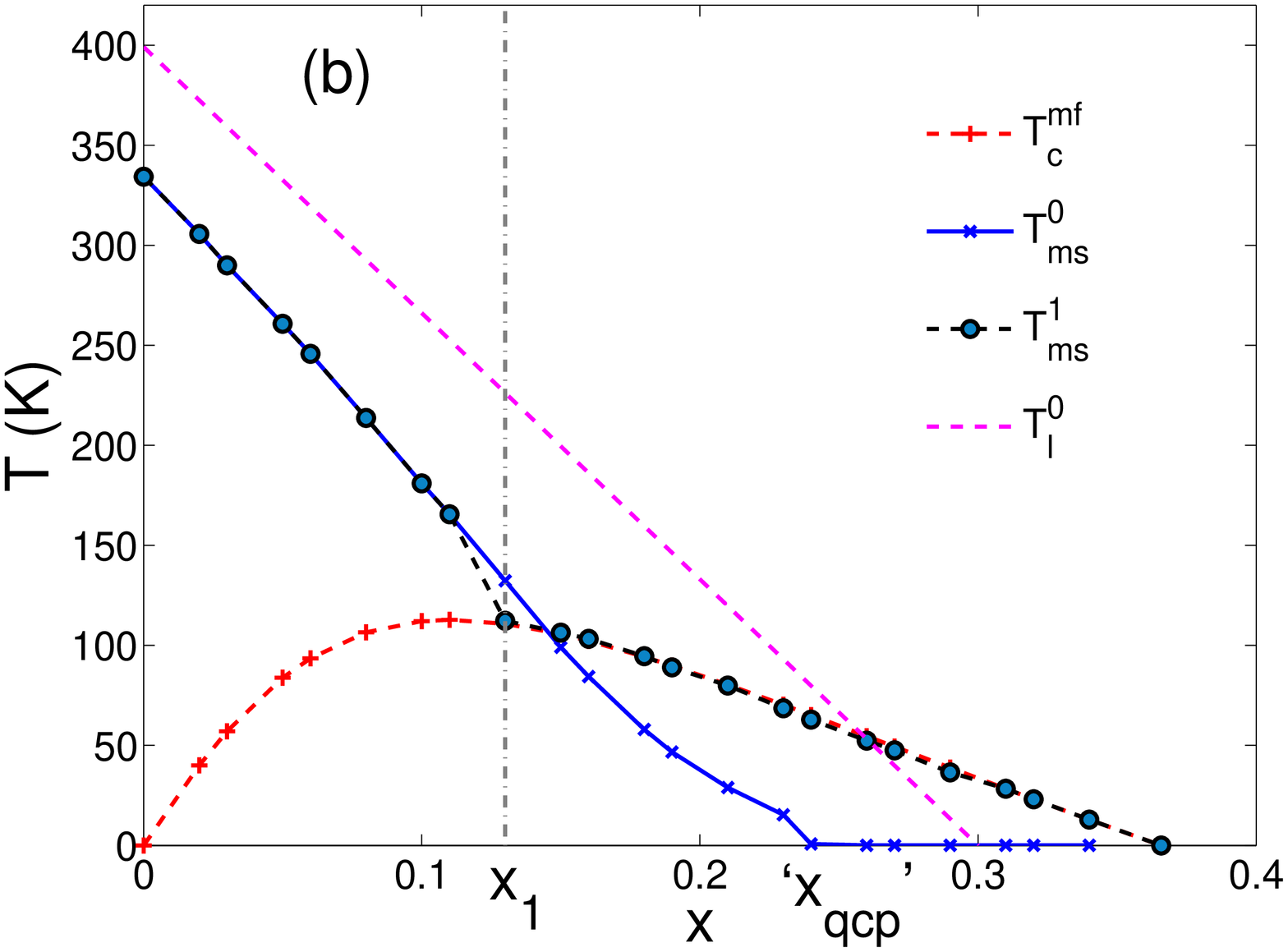}
\end{tabular}
\end{center}
\caption{{\bf (a)} Extraction of $T^1_\mathrm{ms}(x)$ from the positions of the maximum of $|\frac{\partial
\langle \Delta_m \rangle}{\partial T}|\equiv |\partial_T \langle\Delta_m\rangle|$ vs. $T$ curves (upper panel)
at various doping values. 
Two local maxima appear in the underdoped regime, one sharp peak at $T_c$ and a broad maximum at
$T^1_\mathrm{ms}$. $T^1_\mathrm{ms}(x)$ merges with $T_c(x)$ in the overdoped side (inset of upper panel). 
Similar analysis (lower panel) is carried out on $|\frac{\partial \langle\Delta_m\rangle_0}{\partial T}|$ 
(see text for definition) to extract $T^0_\mathrm{ms}$. {\bf (b)} Comparison of $T^*(x)$, identified with
$T^{0,1}_\mathrm{ms}$, with other relevant temperature scales; different pseudogap scenarios
\cite{MRNorman1} are naturally embodied in our results, as discussed in the text.}
\label{fig.PseudogapT}
\end{figure}
 
 Measurements detect a diminution in the density of electron states, one which depends on the direction of
$\mathbf{k}$ along the Fermi surface. Different measurements (e.g. NMR, resistivity, ARPES etc.) show 
characteristic changes at temperatures which differs by 20 K to 40 K \cite{TTimusk}. The `pseudogap temperature' $T^*(x)$ is, therefore, not very well-defined. $T^*$ is generally 
seen to decrease with hole doping $x$, nearly linearly, till it `hits' the $T_c(x)$ curve, around  (but
slightly beyond) $x_\mathrm{opt}$. What happens next is a matter of considerable controversy. Broadly, three
scenarios have been argued for, as described for example in Ref.\onlinecite{MRNorman1}. One of them \cite{NNagaosa} suggests 
that the pseudogap temperature
merges with $T_c(x)$ a little beyond optimum doping. Another scenario \cite{JLTallon,SChakravarty,CMVarma2} is that 
it goes through the $T_c(x)$ dome, reaches zero at a putative quantum critical point $x_\mathrm{qcp}$, which controls the 
universal low temperature behaviour of the cuprate around it in the $(x,T)$ plane. A third \cite{MRNorman1} is that there is 
no $T^*$ beyond the hole concentration $x_1$ at which it `touches' $T_c(x)$. Operationally, we identify the pseudogap 
temperature as one at which the absolute value of the slope of $\bar{\Delta}(x,T)$ as a function of temperature is a local 
maximum, calling it $T_\mathrm{ms}(x)$. In general, this definition leads to two characteristic temperatures. One of them is
at $T_c$ because a part of $\bar{\Delta}(x,T)$ suddenly turns on at $T_c$ due to the onset of global phase coherence, leading to a divergence of the temperature derivative at $T_c$. The other is at a temperature higher than $T_c(x)$ till an $x$ value slightly above
$x_\mathrm{opt}$. This fact leads to two kinds of behaviour for $T_\mathrm{ms}(x)$
(Fig.\ref{fig.PseudogapT}) and thus for the
pseudogap temperature $T^*(x)$ if these two are identified with each other. If we start from the low doping (small $x$)
side, where $T_\mathrm{ms}(x)$ is high and follow it as $x$ increases, noticing its origin in local pairing
and existence even when there is no global order, we see that this branch of $T_\mathrm{ms}(x)$ denoted as 
$T^0_\mathrm{ms}(x)$ in Fig.\ref{fig.PseudogapT} hits the 
$T_c(x)$ line at $x_1$ (Fig.\ref{fig.PseudogapT}(b)), goes through the $T_c$ dome to zero temperature at `$x_\mathrm{qcp}$' 
and continues to be zero thereafter. On the other hand, if beyond $x_1$ we choose the other solution for $T_\mathrm{ms}(x)$
(called $T^1_\mathrm{ms}(x)$ in Fig.\ref{fig.PseudogapT}), which exists because of the long range order
causing `Josephson' or $C$ term in Eq.\eqref{Eq.functional1}, then one has a pseudogap curve which is above
$T_c(x)$ till $x_1$ and is the same as $T_c(x)$ thereafter. These are two of the pseudogap categories 
mentioned above. Different types of experiments are likely to probe different types of pseudogap. For example,
if superconducting phase coherence is destroyed with a magnetic field, so that the $C$ or Josephson term is
ineffective, the observed pseudogap behaviour with $x$ is that of the first category.      
          
At zero temperature the phase coherent classical ground state can be represented in terms of nearest-neighbor 
singlet bond pair fields $\psi_m$ or equivalently $\psi_{i\mu}$ (see Fig.\ref{fig.BondLattice}) as
\begin{subequations} \label{Eq.groundstate_gap}
\begin{eqnarray}
\psi_{ix}&=&-\psi_{jy}=\Delta_0(x)~~~~\forall i,j \label{Eq.groundstate}\\
\Delta_0(x)&=&\Delta_0(0)\left(1-\frac{x}{x_{c_2}}\right)^{\frac{1}{2}}~~~~x\leq x_{c_2}, \nonumber \\
&=&0~~~~x>x_{c_2}. \label{Eq.gap0}
\end{eqnarray}
\end{subequations}
Here, $\Delta_0(x)$ is the zero temperature gap (see Fig.\ref{fig.Tc}), 
$\Delta_0(0)=1/(f\sqrt{b})$ and $x_{c_2}=x_c/(1-2cx_c)$ is obtained from $A(x_{c_2},0)-2C(x_{c_2})=0$.          

\begin{figure}
\begin{center}
\begin{tabular}{c}
\includegraphics[height=6cm]{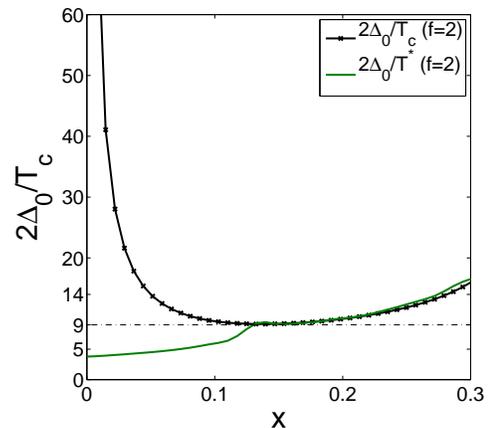}
\end{tabular} 
\end{center}
\caption{ $2\Delta_0(x)/T_c(x)$ and $2\Delta_0(x)/T^*(x)$ as functions of
$x$. Here $T^*(x)$ refers to $T^1_\mathrm{ms}(x)$ (see Fig.\ref{fig.PseudogapT}). The long-dashed line
corresponds to the nearly constant value of $2\Delta_0(x)/T_c(x)$ near optimal doping.}
\label{fig.GaptoTcRatio}
\end{figure}
        
Our choice of the values of $b$ and $f$ fixes the ratio $2\Delta_0/T_0=2/(f\sqrt{b})$ to be around $3-5$, 
which implies that $2\Delta_0(x)/T^*(x)$ also stays close to these values in the underdoped regime 
(Fig.\ref{fig.GaptoTcRatio}). It has been widely reported \cite{MKugler,OFischer} that the ratio of the low temperature 
(`zero temperature') gap to the pseudogap temperature scale, specifically $\Delta_0(x)/T^*(x)$, for a
range of hole doping, especially below the optimum $x$, is about 4.3/2, which is the universal $d$-wave 
BCS value \cite{HWon} for the ratio of zero temperature gap to superconducting transition temperature. Further by choosing $c=0.3$, the ratio $2\Delta_0(x)/T_c(x)$ near optimal doping is see to be  around 10 to 15, as observed in cuprates 
\cite{ADamascelli,OFischer}, being substantially higher than the BCS ratio. In
Fig.\ref{fig.GaptoTcRatio}, the ratio $2\Delta_0(x)/T_c(x)$ is shown to be more or less constant around
optimal doping. The increase of this ratio as $(1-x/x_{c_2})^{-1/2}$ for large $x$ is an
artifact of the chosen classical functional. 

\section{Specific Heat}\label{sec.SpecificHeat}
  
The electronic specific heat of the superconducting cuprates has been measured in many
experiments~\cite{JWLoram1,JWLoram2,JWLoram3}. It consists of a sharp peak near the superconducting
transition temperature $T_c(x)$ and a broad hump around the pseudogap $T^*(x)$ \cite{TMatsuzaki}, both riding on a component
that is clearly linear in $T$ at temperatures $T\geq T^*$  in optimally doped and overdoped samples. Here, we
summarize theoretical results for the specific heat arising from our functional (Eq.\eqref{Eq.functional}),
both with and without magnetic field. A detailed description is given in a separate paper \cite{SBanerjee1}.
The functional captures the thermodynamic probability of (bosonic) Cooper pair fluctuations and yields the
contribution of these fluctuations to the specific heat. Because of our use of a classical functional, the low
temperature behaviour dominated by quantum effects is not properly accounted for; we discuss this below. The low energy 
electronic degree of freedom ignored in our treatment are the fermionic, non-Cooper-pair ones of the degenerate electron gas.  We use the free energy functional (Eq.\eqref{Eq.functional}) to write the specific heat
as
\begin{eqnarray}
&&C_v=\frac{1}{N_b}\frac{\partial\langle\mathcal{F}\rangle}{\partial T}=\frac{1}{N_b}
\left[\frac{1}{T^2}\left(\langle\mathcal{F}^2\rangle-\langle\mathcal{F}\rangle^2\right)\right.\nonumber\\
&+&\left.\frac{\partial A}{\partial
T}\sum_m\left(\langle\Delta_m^2\rangle-\frac{1}{T}\left(\langle\Delta_m^2\mathcal{F}\rangle-\langle\Delta_m^2\rangle
\langle\mathcal{F}\rangle\right)\right)\right]\nonumber\\
&& \label{Eq.specificheat}
\end{eqnarray}
where $\frac{\partial A}{\partial T}=(f^2\exp(T/T_p)+A/T_p)$ for the particular choice of $A$ as in
Eq.\eqref{Eq.A}. Clearly the second term in Eq.\eqref{Eq.specificheat} arises from the fact that $\mathcal{F}$
is an effective low energy functional whose basic parameters, e.g. $A$, can be temperature dependent. We
evaluate $C_v$ from Eq.\eqref{Eq.specificheat} for different values of doping $x$ and temperature $T$ by MC
sampling of finite 2D systems as mentioned in Section \ref{sec.SuperfluidDensity}. The simulations have
been carried out with $f=2$ (see Section \ref{subsec.GLParameters}) while choosing $\Delta_0(x=0)\simeq 54$
meV, so that $T_0=400$ K.

\begin{figure}
\begin{center}
\begin{tabular}{c}
\includegraphics[height=6cm]{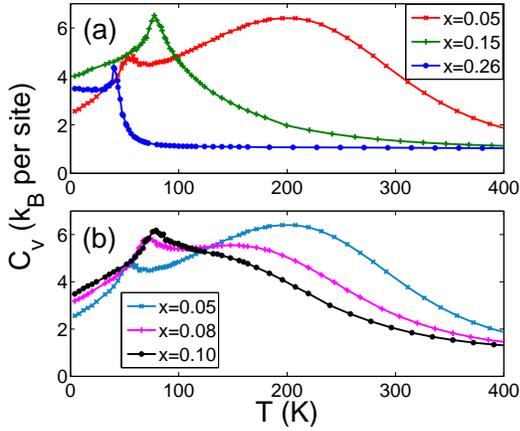}
\end{tabular} 
\end{center}
\caption{ {\bf (a)} Specific heat obtained from MC simulation of our model
(Eq.\eqref{Eq.functional}). Panel {\bf (b)} shows the evolution of the broad maximum around $T^*$ with doping in the
underdoped region.}
\label{fig.SpecificHeat}
\end{figure}

We notice that in both theory (see Fig.\ref{fig.SpecificHeat}) and experiment\cite{JWLoram3,PCurty,CPMoca},
there is a sharp peak in $C_v$ around $T_c$ (or $T_\mathrm{BKT}$ in our case to be more precise). The peak
amplitude increases as $x$ increases, leading to a BCS like shape in the overdoped side. In addition, there is
a hump \cite{TMatsuzaki}, relatively broad in temperature, centered around $T^*$. The hump is most clearly visible in the
calculation for the underdoped regime where $T^*$ and $T_c$ are well separated; its size in the theory depends
on $A$ and $B$ (Eq.\eqref{Eq.GLparameters}). In experiments, for the
underdoped side, its beginnings can be seen; unfortunately there are very few experiments over a wide enough
temperature range to encompass the hump fully in this doping regime. The two features, namely the peak and the
hump, and their evolution with $x$ can be rationalized physically. The peak is due the low-energy pairing
degrees of freedom which cause long-range phase coherence leading to superconductivity; these are phase
fluctuations in the underdoped regime. The hump is mainly associated with the regime where the energy
associated with order parameter magnitude fluctuations changes rapidly with temperature. Since this change
is a crossover centered around $T^*$ rather than a phase transition, there is only a specific heat hump, not a
sharp peak or discontinuity. For small $x$, $T^*>>T_c$ and so we see that the hump is well-separated from the
peak. As $x$ increases, $T^*$ approaches $T_c$, and in the overdoped regime, these are not separated, and
there is no hump, only a peak corresponding to the superconducting transition.

\begin{figure}
\begin{center}
\begin{tabular}{c}
\includegraphics[height=6cm]{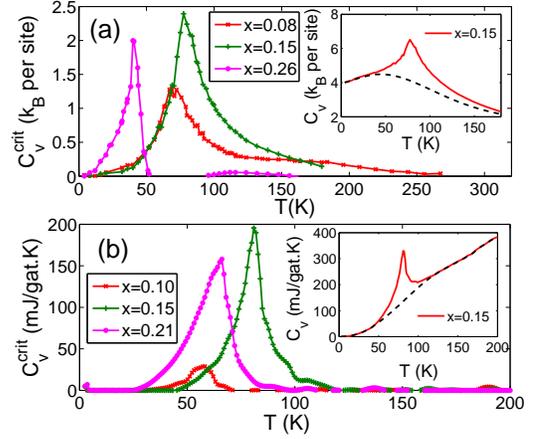}
\end{tabular} 
\end{center}
\caption{ {\bf (a)} The `critical' peak appearing near $T_c$ for three values of $x$. The inset demonstrates the
procedure used for the subtraction of the `non-critical' background (dashed line), as mentioned in the text. {\bf (b)}
Analogous plot for the experimental specific heat data for
$\mathrm{Y_{0.8}Ca_{0.2}Ba_2Cu_3O_{7-\delta}}$ from \cite{JWLoram2}. Here, $x$ values are estimated using the
empirical form of Persland et al. \cite{MPersland}. Again, the inset shows the subtracted background (dashed line) for $x=0.15$.}
\label{fig.SpecificHeatCritPeak}
\end{figure}

\begin{figure}
\begin{center}
\begin{tabular}{c}
\includegraphics[height=6cm]{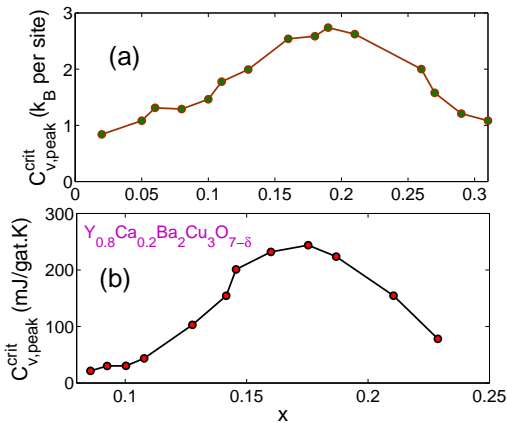}
\end{tabular} 
\end{center}
\caption{ {\bf (a)} Evolution of the height of the specific heat peak appearing near $T_c$ with doping,
compared with the analogous plot {\bf (b)} obtained from experimental data for
$\mathrm{Y_{0.8}Ca_{0.2}Ba_2Cu_3O_{7-\delta}}$ \cite{JWLoram2}.}
\label{fig.SpecificHeatCritPeakHeight}
\end{figure}

In order to compare our results with experiments, in particular the features related to critical fluctuations near $T_c$),  
we remove the contributions that are special to the chosen
classical functional and are not connected with the Cooper-pair degrees of freedom in the real systems. Firstly, at low
temperatures, $T<<T_c$, the fact that we have a classical functional here leads to a large specific heat of
the order of the Dulong-Petit value and there is an additional contribution ($\propto \frac{\partial
A}{\partial T}$, see Eq.\eqref{Eq.specificheat}) due to temperature dependence of $A$, whereas the actual specific heat is 
expected to be small because of quantum effects (it is $\sim T^2$ due to nodal quasiparticles \cite{NMomono1}). To account 
for this difference, we compute the leading low-temperature contribution to the specific heat arising from our 
functional (Eq.\eqref{Eq.functional}). Similarly at high temperatures $T>T^*$, the contribution from pairing
degrees of freedom for the actual system is expected to be small, whereas from the functional
(Eq.\eqref{Eq.functional}) it is not so due to the simplified from used for the single-site term
(Eq.\eqref{Eq.functional0}). We compute $C_v$ from a high temperature expansion for the
intersite term in Eq.\eqref{Eq.functional}. We interpolate for the specific heat using the low and high
temperature expansion results, and subtract the resulting part (includes the hump) from the calculated specific heat. 
This subtracted specific heat is plotted in Fig.\ref{fig.SpecificHeatCritPeak}(a) for three values of doping. These are 
compared with the experimental electronic specific heat data of Ref.~\cite{JWLoram2} for YBCO after analogous subtraction of 
a `non-critical' smooth part obtained from interpolation between low and high temperature regions (excluding the peak) is 
done (see inset of Fig.\ref{fig.SpecificHeatCritPeak}(b)). This procedure also removes linear $T$ contribution to
specific heat arising from unpaired low energy electronic degrees of freedom present in the system but not in
our functional (Eq.\eqref{Eq.functional}). Since the peaks are large and occur over a narrow temperature
near $T_c$, they are relatively free from possible errors due to the subtraction procedure mentioned above.
The experimental and theoretical results for specific heat peaks are shown separately in 
Fig.\ref{fig.SpecificHeatCritPeak}. We see that they compare well with each other. The qualitative agreement is brought 
out clearly in Fig.\ref{fig.SpecificHeatCritPeakHeight} where we plot the specific heat peak height with $x$ and compare 
the dependence with what is observed in experiment. This implies that our model for the bond pairs and
their interaction to generate a $d$-wave superconductor is a faithful representation of the relevant
superconductivity related degrees of freedom. 

\begin{figure}
\begin{center}
\begin{tabular}{c}
\includegraphics[height=6cm]{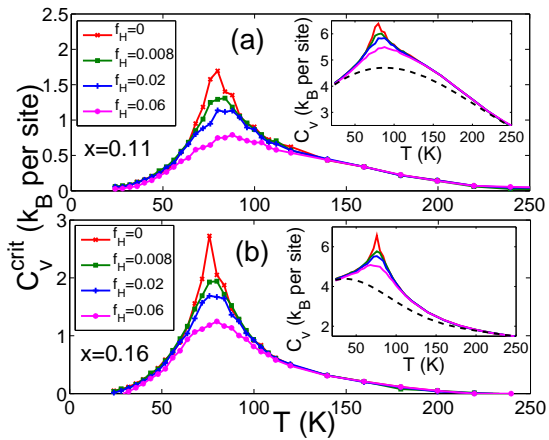}
\end{tabular} 
\end{center}
\caption{ Effect of a magnetic field on the specific heat peak for {\bf (a)} x=0.11 and {\bf (b)} x=0.16.
The subtraction procedure employed in Fig.\ref{fig.SpecificHeatCritPeak} is used here as well, as shown in the insets.}
\label{fig.SpecificHeatPeakField}
\end{figure}

The effects of a magnetic field on the specific heat have been cataloged in \cite{AJunod,HWen} where it is
found that the specific heat peak near $T_c$ is increasingly smoothed out with magnetic field, but the
peak position does not shift by much, especially in highly anisotropic systems such as Bi2212 and Bi2201. This
effect is most clearly visible for small $x$, and occurs even for magnetic fields as small as a few Tesla. We
assume that only the intersite term depends on the vector potential
$\mathbf{A}$, via the Peierls phase factor, namely that $(\phi_m-\phi_n)$ in Eq.\eqref{Eq.functional1} is
replaced by $(\phi_m-\phi_n-\frac{2e}{\hbar c}\int_{\mathbf{R}_m}^{\mathbf{R}_n}\mathbf{A}.d\mathbf{l})$. The
resulting specific heat `peak' curves obtained from MC simulations are plotted in
Fig.\ref{fig.SpecificHeatPeakField} for two $x$ values at different values of $f_H=Hl^2/\Phi_0$ i.e. the flux
going through each elementary plaquette of the bond lattice in units of the fundamental flux quantum
$\Phi_0=hc/(2e)$, where $\mathbf{H}$ is the applied uniform magnetic field perpendicular to the plane (i.e.
$\mathbf{H}=H\hat{z}$) and we assume the extreme type-II limit. The results compare well with those of experiment \cite{HWen}.      

\section{Vortex Structure and Energetics} \label{sec.Vortex}

\begin{figure}
\begin{center}
\begin{tabular}{ccc}
\includegraphics[height=3.8cm]{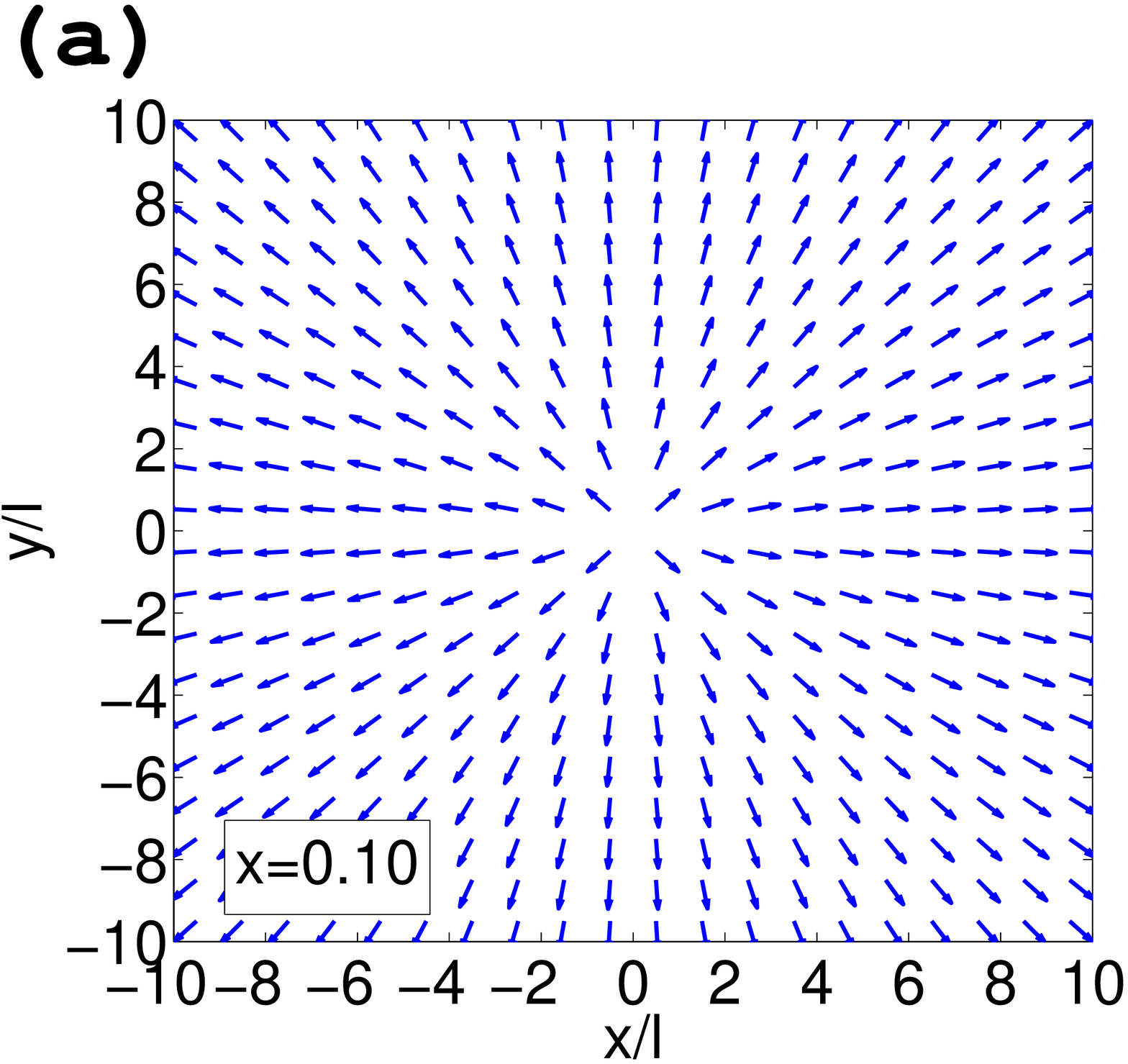}& 
\includegraphics[height=3.8cm]{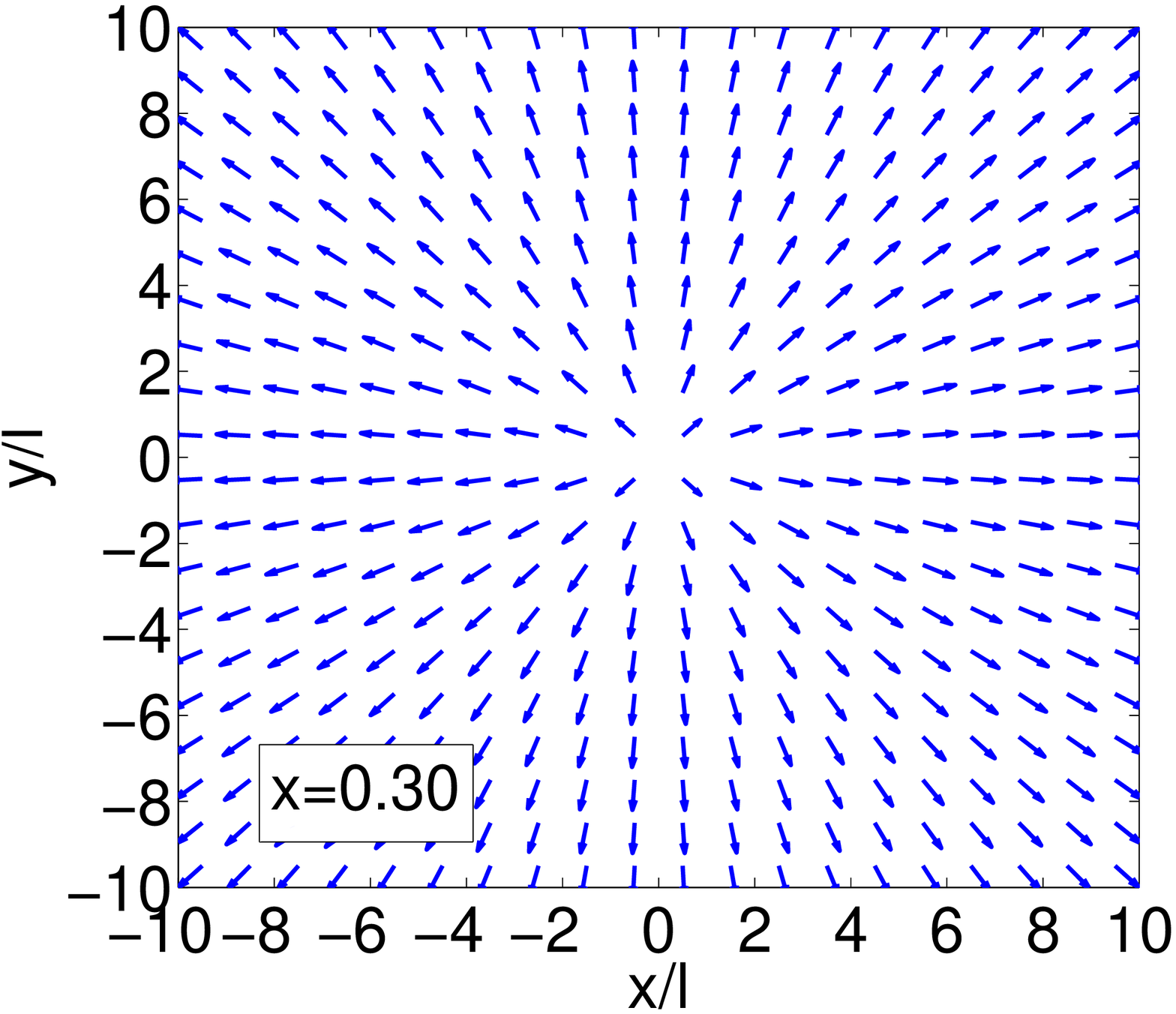}\\
\includegraphics[height=3.8cm]{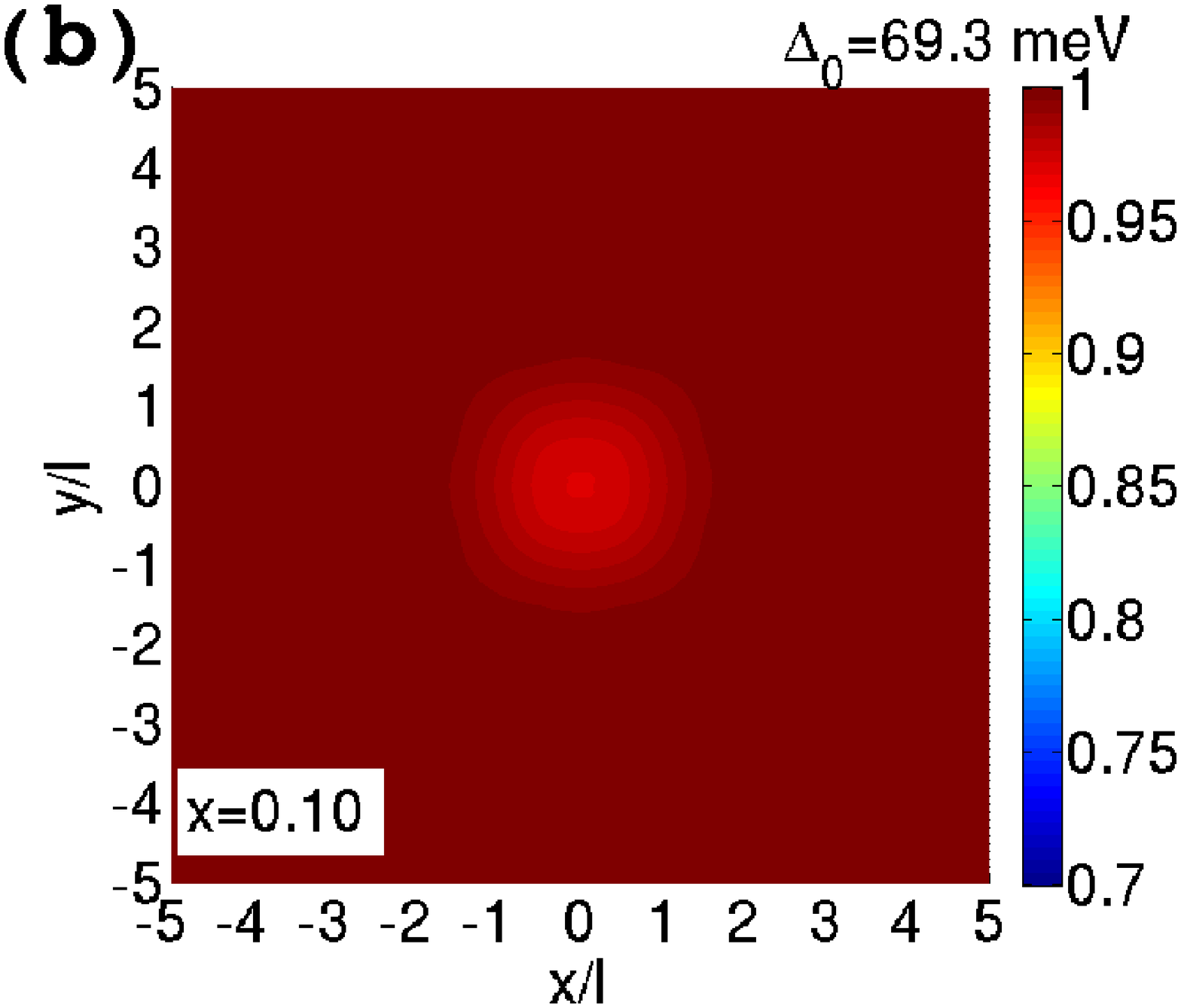}& 
\includegraphics[height=3.8cm]{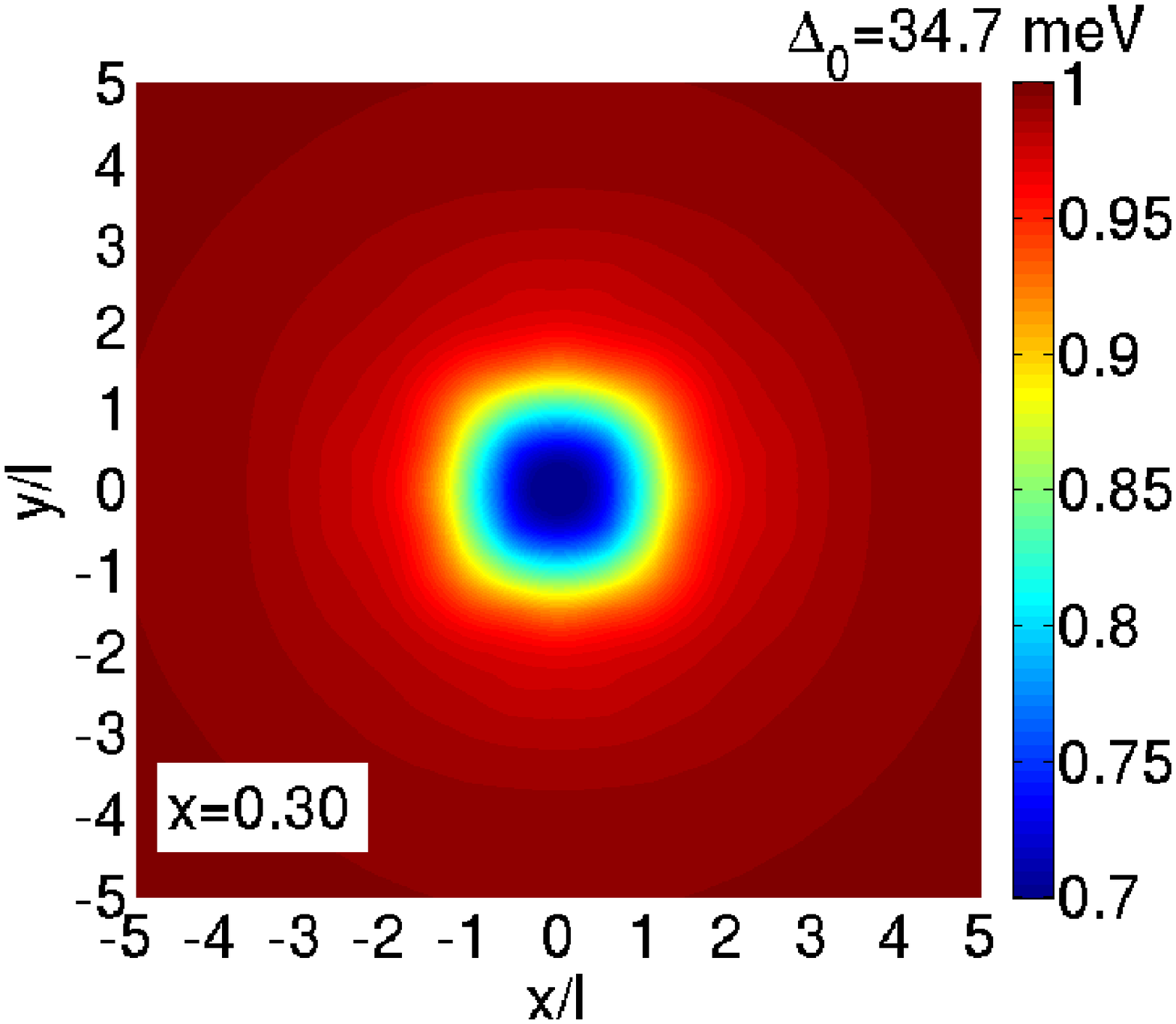}
\end{tabular}
\includegraphics[height=6cm]{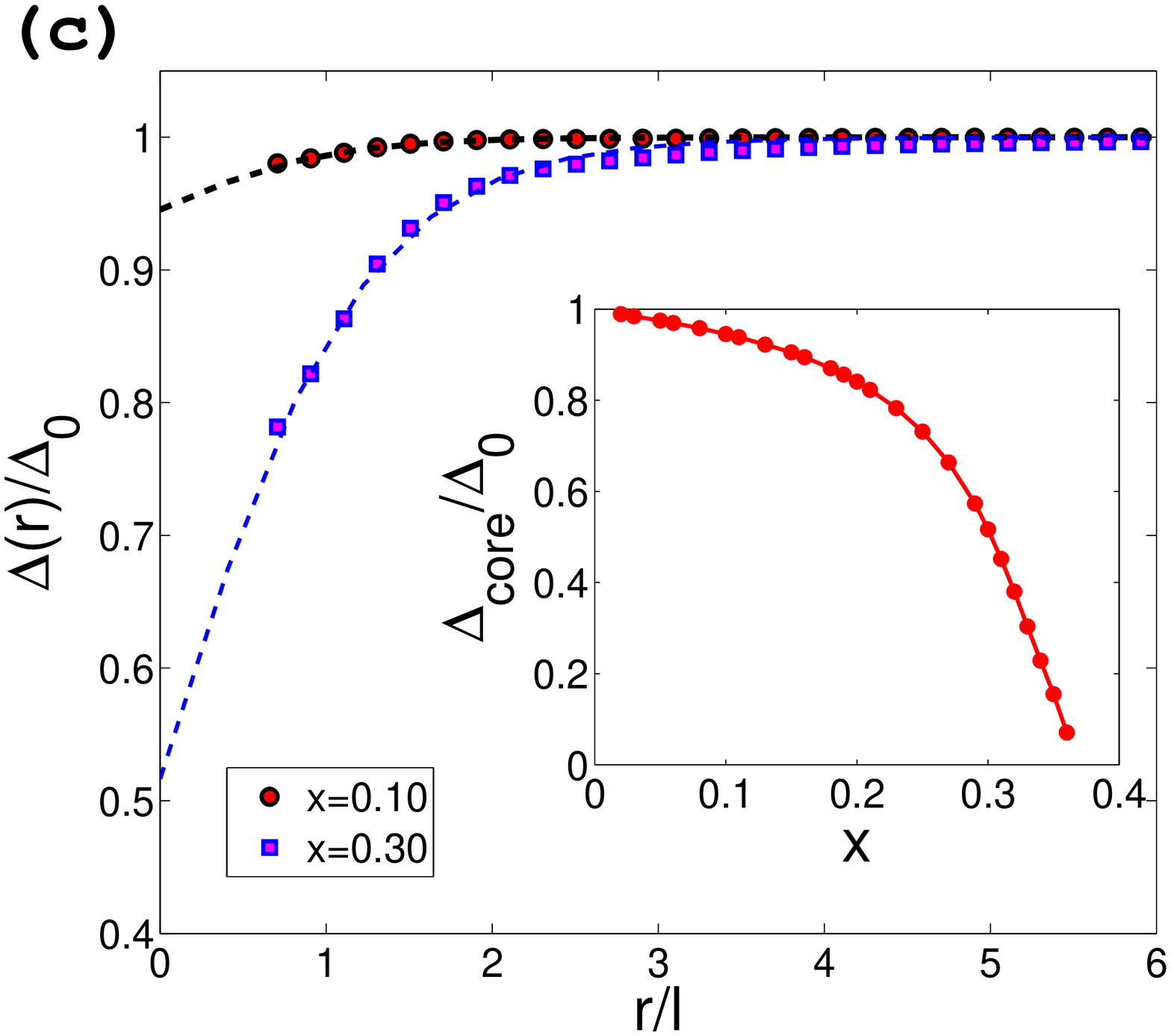}\\
\end{center}
\caption{{\bf (a)} Single vortex configuration for $x=0.10$ and $x=0.30$. Arrows indicate the equivalent
planar spins. A sublattice transformation has been performed on the
phases for convenience of representation. {\bf (b)} Variation of the magnitude of the bond pair field near the
vortex core for the aforementioned values of $x$. The magnitude is plotted in units of its maximum value
attained in the bulk, $\Delta_0$ (mentioned at the top of each color bar). {\bf (c)} The angular averaged gap
magnitude $\Delta(r)$ (normalized by $\Delta_0$) as function of distance from the core for the two $x$ values.
Inset shows the doping dependence of the magnitude at the core, $\Delta_\mathrm{core}$, estimated by fitting 
$\Delta(r)$ with $\Delta_0\tanh{(r/\xi_c)}+\Delta_\mathrm{core}$, while $\xi_c$ and $\Delta_\mathrm{core}$ are kept as 
fitting parameters.}
\label{fig.VortexConfiguration}
\end{figure}

We use the functional (Eq.\eqref{Eq.functional}) to find the properties of vortices that 
are topological defects in the ordered phase. This has been extensively done in the GL theory for conventional
superconductors \cite{MTinkham}. We use the free energy functional of Eq.\eqref{Eq.functional} at $T=0$, where 
it describes the ground state properties, to generate a single vortex configuration by minimizing $\mathcal{F}$
with respect to $\Delta_m$ and $\phi_m$ at each site while keeping the topological constraint of total $2\pi$
winding of the phase variables at the boundary of a $N_b\times N_b$ lattice. This is a standard way of
generating a stable single $k=1$ vortex configuration with the vortex core at the middle of the central square
plaquette in the computational lattice. The results for $\{\Delta_m,\phi_m\}$ are shown in
Fig.\ref{fig.VortexConfiguration} for two different values of hole doping $x$, namely $x=0.10$ (underdoping)
and $x=0.30$ (overdoping). Fig.\ref{fig.VortexConfiguration}(a) shows the order parameter at a point $m$ on
the square lattice as an arrow whose length is proportional to the value of $\Delta_m$ there, and whose
inclination to the $x$-axis is equal to the phase angle $\phi_m$.

\begin{figure}
\begin{center}
\begin{tabular}{c}
\includegraphics[height=5cm]{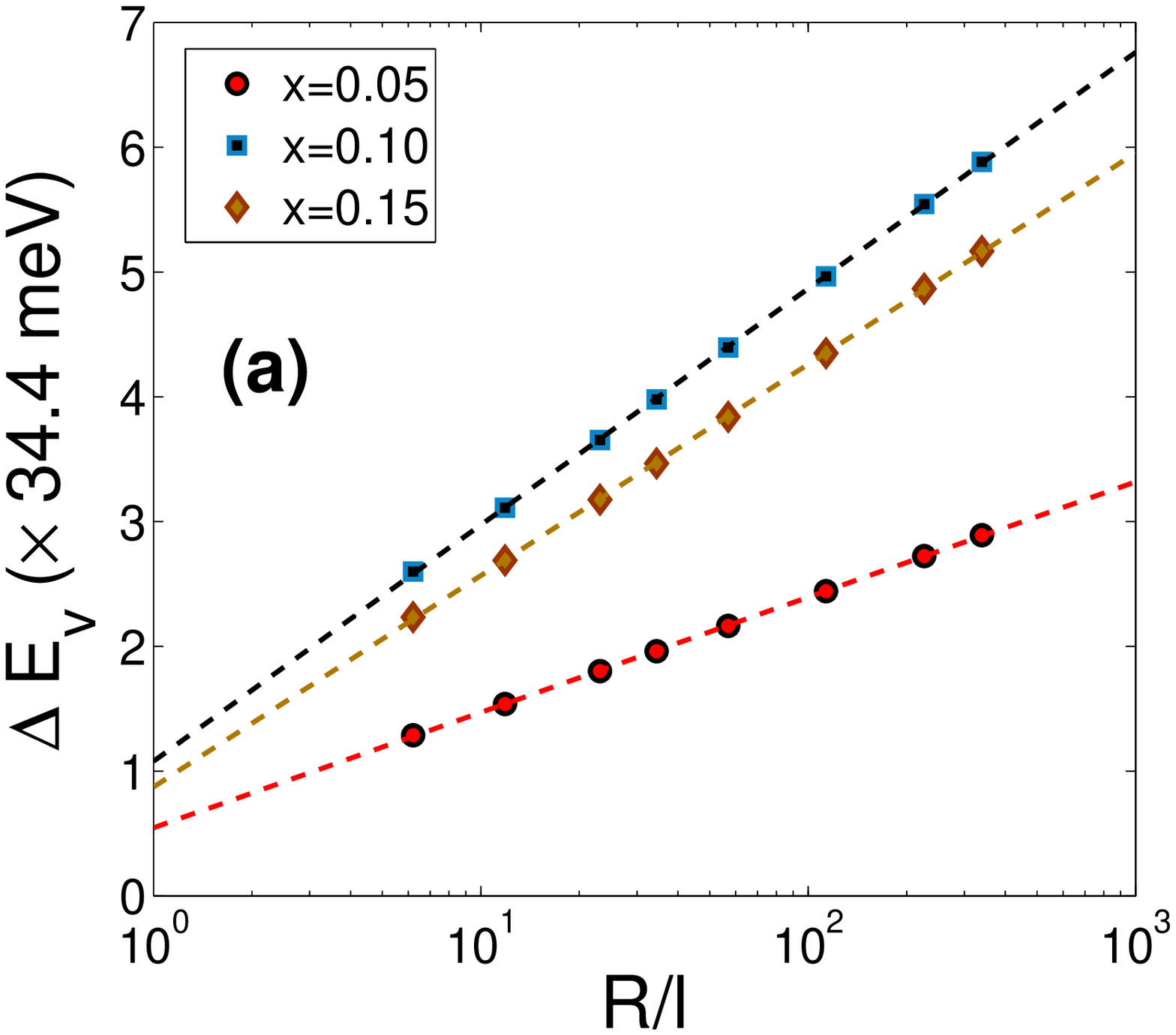}\\ 
\includegraphics[height=5cm]{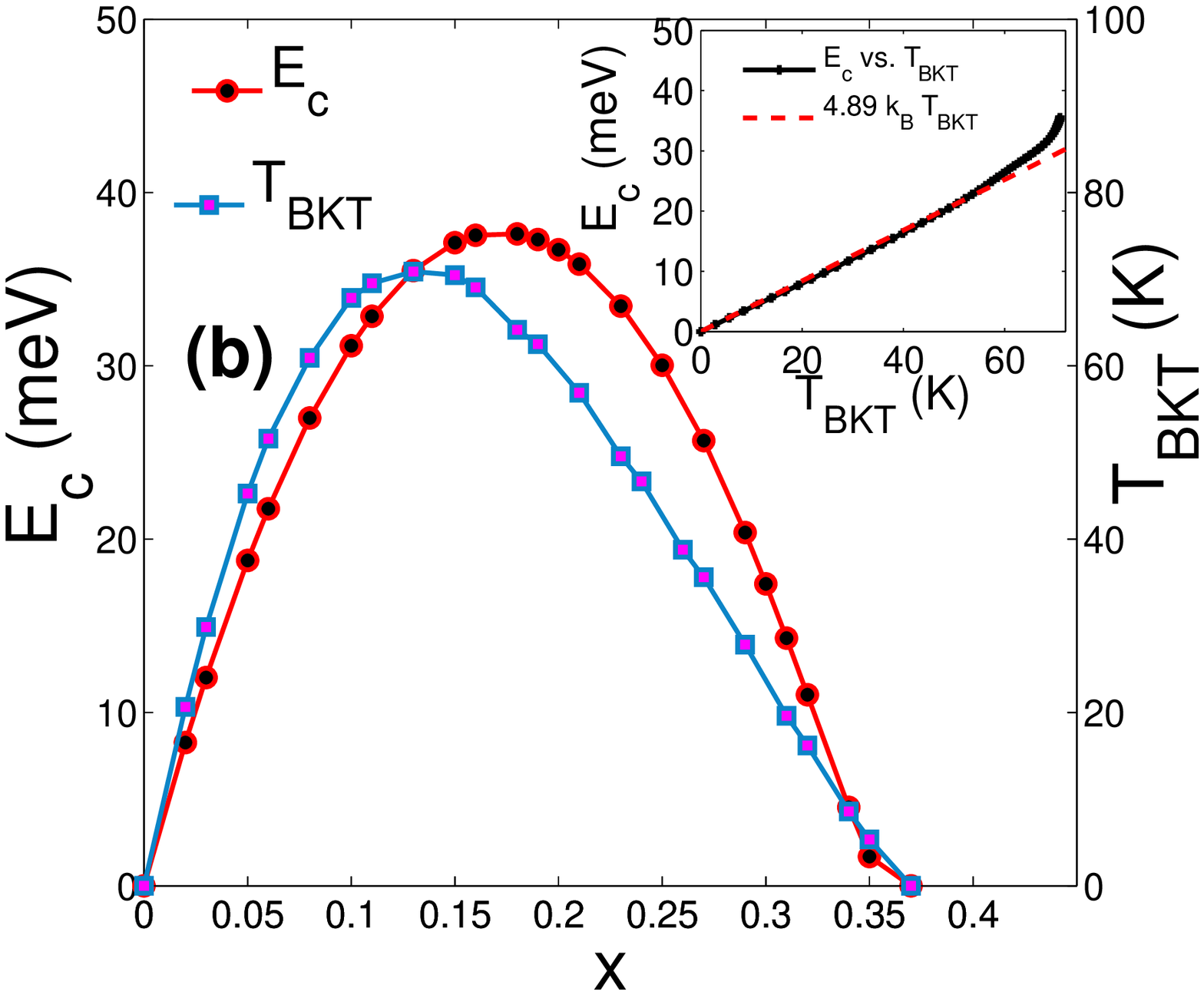}
\end{tabular}
\end{center}
\caption{{\bf (a)} The excess energy of a vortex $\Delta E_v$ as a function of system size (see main text) for
three values of $x$. Intercepts of the dashed lines with the vertical axis yield the values of 
the corresponding core energies $E_c$. {\bf (b)}
$E_c$ is compared with $T_c$. Like $\rho_s(0)$ (see Fig.\ref{fig.SuperfluidDensity}(b)), $E_c$ peaks
at $x\simeq 0.19$. The inset shows the proportionality of $E_c$ and $T_\mathrm{BKT}$ in the underdoped side.}
\label{fig.VortexCoreEnergy}
\end{figure}

We notice that for the underdoped cuprate (e.g. $x=0.10$) unlike the overdoped one ($x=0.30$), the order parameter magnitude 
does not decrease by much as one moves radially inwards from far to the core (Fig.\ref{fig.VortexConfiguration}(b)). This is characteristic of a phase or Josephson vortex
whose properties have been investigated for coupled Josephson junction lattice system \cite{CJLobb}. We
propose therefore that vortices in cuprates in the underdoped regime are essentially
Josephson vortices. This is natural here because the Cooper pair amplitude $\Delta_m$ has sizable
fluctuations only close to $T^*$ which is well separated from $T_c$ ($T_c<<T^*$) in the underdoped regime so
that near $T=0$, there are very small $\Delta$ fluctuations. Further, for a lattice system (and not for a
strict continuum) such a defect is topologically stable since the smallest possible perimeter is the elementary
square. On the other hand, beyond optimum doping where, according to
Fig.\ref{fig.PseudogapT}, $T^*$ coincides with $T_c$, the order parameter magnitude $\Delta_m$ decreases
substantially on moving radially inwards towards the vortex core, very much like a `conventional'
superconducting or BCS vortex. The variation of the normalized magnitude of the bond pair field
$\Delta(r)/\Delta_0$ with the radial distance $r$ from the vortex core in
the two cases is shown in detail in Fig.\ref{fig.VortexConfiguration}(c), which clearly illustrates the
difference between the behavior in the two cases. The inset of Fig.\ref{fig.VortexConfiguration}(c) shows the
extrapolated values of the magnitudes ($\Delta_\mathrm{core}$) at the core ($r=0$) as a function of
$x$, indicating that there is a smooth crossover from a Josephson-like vortex to a BCS-like vortex
with increasing hole density $x$.
  
The core energy $E_c$ of a single vortex is naturally described as the extra energy $\Delta E_v=E_v-E_0$
where $E_0$ is the energy of the ground state configuration (the $\mathrm{Ne'el}$ ordered state in this case) and
$E_v$ is the total energy of a single vortex configuration, from which the elastic energy due to phase
deformation \cite{PMChaikin} is subtracted, i.e.
 \begin{eqnarray}
\Delta E_v&=&E_c+\pi \rho_s(0) \ln(R/l)
\end{eqnarray}
The quantity $R$ is defined as $R=(N_b-1)l/\sqrt{\pi}$, where $l$ is the lattice constant
of the bond lattice, so that $\pi R^2$ is the area of the computational lattice. 
We plot in Fig.\ref{fig.VortexCoreEnergy}(b) the core energy $E_c$ as a function of $x$, both its absolute
value and its ratio with $T_c$. $E_c$ has been estimated from the intercept of the $\Delta E_v$ vs. $\ln(R/l)$
(different system sizes) straight line with the energy axis. We notice that for small $x$,
$E_c(x)\propto T_c(x)$ (inset of Fig.\ref{fig.VortexCoreEnergy})(b), not surprising from XY model
considerations \cite{LBenfatto}.

\section{Electron Spectral Function and ARPES} \label{sec.SpectralDensity}

The cuprate superconductor obviously has both electrons, and Cooper pairs of the {\em same} electrons,
coexisting with each other. In a GL-like approach such as ours, only the latter are explicit, while the
former are `integrated out'. However, effects connected with the pair degrees of freedom are explored
experimentally via their coupling to electrons, a very prominent example being photoemission in which the
momentum and energy spectrum of electrons ejected from the metal by photons of known energy and momentum is
investigated. Since ARPES (angle resolved photoemission spectroscopy) \cite{ADamascelli,JCCampuzano} is a
major and increasingly high-resolution \cite{JDKoralek} source of information from which the behaviour of pair
degrees of freedom is inferred, we mention here some experimental consequences of a theory of the coupling
between electrons and the complex bond pair amplitude $\psi_m$. The theory as well as a number of its
predictions (in agreement with ARPES measurements) are described in detail in Ref.\onlinecite{SBanerjee3}.

In formulating a theory of the above kind, one faces the difficulty of having to develop a description of
electrons in a presumably strongly correlated system such as a cuprate, which is viewed as a doped Mott
insulator \cite{PALee} with strong  low-energy antiferromagnetic correlation between electrons at nearest
neighbor sites \cite{MAKastner}. In particular, one needs to commit oneself to some model for electron dynamics which then 
implies an approach to the coupling between electronic and pair degrees of freedom. We develop what we
believe is a minimal theory, appropriate for low-energy physics. We assume that for low energies $|\omega|\leq
\Delta_0$, well-defined electronic (tight-binding lattice) states with renormalized hopping amplitudes
$t,~t',~t''$ etc. exist and couple to low-energy pair fluctuations $\psi_m=\psi_{i\mu}=\langle (a_{i\downarrow}
a_{i+\mu\uparrow}-a_{i\uparrow} a_{i+\mu \downarrow})/2\rangle$ (see Fig.\ref{fig.BondLattice}).
Superconducting order (more precisely, phase stiffness) and fluctuations in it are reflected respectively in
the average $\langle \psi_{i\mu}(\tau)\rangle$  and the correlation function $\langle
\psi_{i\mu}(\tau)\psi^*_{j\mu'}(\tau')\rangle$ (or its Fourier transform $D_{\mu\mu'}(2\mathbf{q},iz_m)$,
$z_m=2m\pi/\beta$ being the bosonic Matsubara frequency where $m$ is an integer). A nonzero value of 
$\langle \psi_{i\mu}(\tau)\rangle$ in the `AF' long-range ordered phase below $T_c$ leads to the well known 
Gor'kov $d$-wave Green's function and quasiparticles with spectral gap 
$\Delta_\mathbf{k}=(\Delta_d/2)(\cos{k_xa}-\cos{k_ya})$. The correlation function $D_{\mu\mu'}(\mathbf{q},\omega)$ has a 
generic form for small $q$ and $\omega$ which can be related to the functional (Eq.\eqref{Eq.functional}). 

The coupling between low excitation energy electrons and low-lying pair fluctuations (both inevitable) leads to a self 
energy with a significant structure as a function of electron momentum $\mathbf{k}$ and excitation energy $\omega$.
 Physically, we have electrons (e.g. those with energy near the Fermi energy) moving in a medium of pairs which
have finite range `AF' or $d$-wave correlation for $T>T_c$ and have long-range order of this kind for $T<T_c$
(in addition to `spin wave' like fluctuations). The electrons exist both as constituents of Cooper pairs and as
individual entities; the pairs and the electrons are in mutual `chemical' equilibrium. The energy shift or
dynamic polarization of electrons due to this process leads to a number of effects which are described in
\cite{SBanerjee3}. For example, for $T>T_c$ we find a pseudogap in electronic density of states which
persists till $T^*$. We get Fermi arcs \cite{MRNorman2,ADamascelli,JCCampuzano} i.e. regions  on the putative
Fermi surface where the quasiparticle spectral density has a peak at zero excitation energy in contrast to the
pseudogap region where the peak is not at the Fermi energy. The antinodal pseudogap `fills up' between $T_c$ and $T^*$ with 
increasing temperature. Below $T_c$, there is a sharp antinodal quasiparticle peak whose strength is related to the 
superfluid density  as observed in experiment \cite{DLFeng}. We also obtain a `bending' or departure of the 
$\Delta_\mathbf{k}$ vs. $\mathbf{k}$ curve from the mean-field canonical $d$-wave form due to order parameter or 
`spin wave' fluctuations. Here we only outline our theoretical approach and show how a temperature $T^\mathrm{an}$ can be 
obtained from the filling in of the antinodal pseudogap above $T_c$. We find that $T^\mathrm{an}$ compares well in its 
magnitude and $x$-dependence with other measures of the pseudogap temperature scale described in Section \ref{sec.Gap}.   

The physical quantity of interest is 
\begin{eqnarray}
\mathcal{A}(\mathbf{k},\omega)&=& -\frac{2}{\pi} \mathrm{Im} [G(\mathbf{k},i\nu_n\rightarrow\omega+i\delta)
\end{eqnarray}
(the fermionic Matsubara frequency, $\nu_n=(2n+1)\pi/\beta$, $n$ being an integer). Assuming translational
invariance one has the Dyson equation for $G$, namely
\begin{eqnarray}
G^{-1}(\mathbf{k},i\nu_n)&=&(G^0)^{-1}(\mathbf{k},i\nu_n)-\Sigma(\mathbf{k},i\nu_n) \label{Eq.DysonEquation}
\end{eqnarray}
where $\Sigma(\mathbf{k},i\nu_n)$ is the self energy.

$G^0(\mathbf{k},i\nu_n)$ is described in terms of a spectral density in the 
usual Lehmann representation \cite{GDMahan}. The spectral density for low excitation energies has a Dirac $\delta$-function 
part i.e. $\mathcal{A}^0(\mathbf{k},\omega)=z_\mathbf{k}\delta(\omega-\xi_\mathbf{k})$ where $\xi_\mathbf{k}$
is the effective quasiparticle energy measured from the chemical potential $\mu$ and $z_\mathbf{k}$ ($<1$) 
is the quasiparticle residue. In the `plain vanilla' or renormalized tight-binding free-particle theory 
\cite{PWAnderson3,BEdegger} $z_\mathbf{k}=1$ and $\xi_\mathbf{k}=\epsilon^\mathrm{eff}_\mathbf{k}-\mu$ with
$\epsilon_\mathbf{k}^\mathrm{eff}=g_t\sum_{(\mathbf{R}_i-\mathbf{R}_j)}t_{ij}\exp[-i\mathbf{k}.(\mathbf{R}_i-\mathbf{R}_j)]$,
so that $G^0(\mathbf{k},i\nu_n)=1/(i\nu_n-\xi_\mathbf{k})$. The factor $g_t$ is due to correlation effects  calculated in 
the Gutzwiller approximation \cite{BEdegger} which projects out states with doubly occupied sites; one further assumes that 
the renormalized quasiparticles propagate coherently. 

\begin{figure}
\begin{center}
\begin{tabular}{c}
\includegraphics[height=3.5cm]{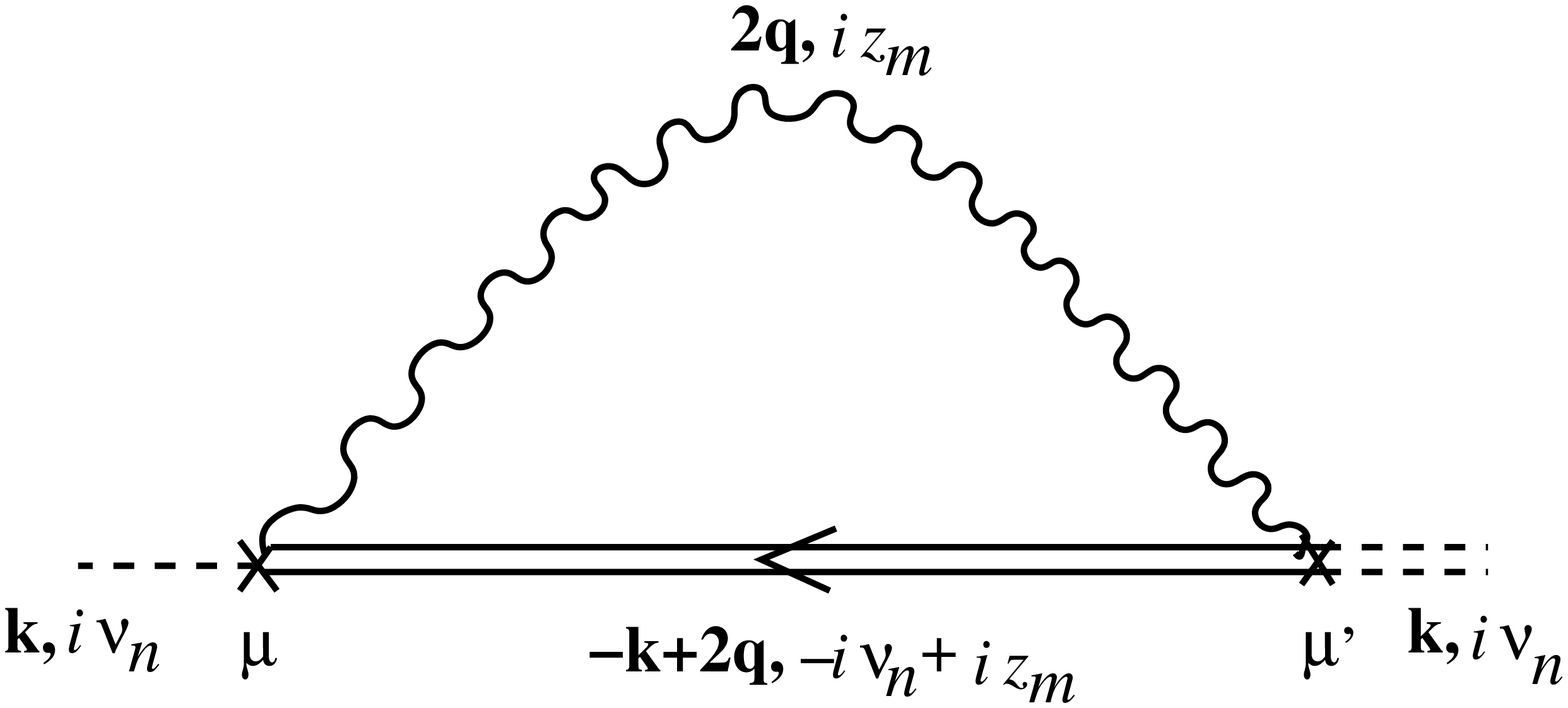}
\end{tabular}
\end{center}
\caption{Self energy approximation used to calculate the electron Green function $G(\mathbf{k},i\nu_n)$. The
wavy line denotes the pair propagator $D_{\mu\mu'}(2\mathbf{q},iz_m)$ and the line with an
arrowhead pointing towards left indicates the full electron Green function
$G(-\mathbf{k}+2\mathbf{q},-i\nu_n+iz_m)$ (see text). The external lines (dashed) at two ends of the diagram represent bare
(left) and true (right) electron propagators. In the static
approximation $D_{\mu\mu'}(2\mathbf{q},iz_m)\equiv (D_{\mu\mu'}(2\mathbf{q})/T^2)\delta_{z_m,0}$ and the
summation over the internal bosonic Matsubara frequency in the above diagram drops out (see Appendix \ref{appendix.SelfEnergy}).}
\label{fig.SelfEnergyApproximation}
\end{figure}

We use a standard approximation for $\Sigma(\mathbf{k},i\nu_n)$ which is shown diagrammatically in
Fig.\ref{fig.SelfEnergyApproximation}. This describes a `phonon' like process neglecting vertex corrections; 
the propagating electron become a Cooper pair (boson) plus an electron in the intermediate state; these recombine to 
give a final state electron with the same $(\mathbf{k},i\nu_n)$. The internal propagator in 
Fig.\ref{fig.SelfEnergyApproximation} is the true or full propagator $G$. However, in common with
general practice, we find $\Sigma$ and thence $G$ by inserting $G^0$ instead of $G$ in the former.
This is known to be quite accurate \cite{GDMahan}, e.g. for the coupled electron-phonon system.

In the static approximation valid at high temperatures when the pair lifetime $\tau_p>>1/(k_BT)$ 
(see Appendix \ref{appendix.SelfEnergy}), the general algebraic expression for 
$\Sigma(\mathbf{k},i\nu_n)$ is
\begin{eqnarray}
&&\Sigma(\mathbf{k},i\nu_n)=\nonumber \\
&& -\frac{1}{N}\sum_{\mathbf{q},\mu,\mu'} G^0(-{\mathbf{k}+2\mathbf{q}},-i\nu_n)
D_{\mu\mu'}(2\mathbf{q})f_\mu(\mathbf{k},\mathbf{q})f_{\mu'}(\mathbf{k},\mathbf{q})\nonumber \\
\label{Eq.StaticSelfEnergy}
\end{eqnarray}   
where $N$ is the total number of Cu sites on a single $\mathrm{CuO_2}$ plane and $\mu$, $\mu'$ refer to the direction of the bond i.e. $x$ or $y$. The static pair propagator is 
$D_{\mu\mu'}(2\mathbf{q})=T^2D_{\mu\mu'}(2\mathbf{q},0)$ (see Fig.\ref{fig.SelfEnergyApproximation}) where
$D_{\mu\mu'}(2\mathbf{q})=\sum_\mathbf{R} D_{\mu\mu'}(\mathbf{R}) \exp{(-i2\mathbf{q}.\mathbf{R})}$ with
$D_{\mu\mu'}(\mathbf{R})=\langle \psi_\mu(\mathbf{R})\psi^*_{\mu'}(\mathbf{0})\rangle$. Since the XY-like interaction term 
(Eq.\eqref{Eq.functional1}) between nearest-neighbor bond pairs (see Fig.\ref{fig.BondLattice}) is antiferromagnetic,
\begin{eqnarray}
D_{xx}(\mathbf{R})=D_{yy}(\mathbf{R})=-D_{xy}(\mathbf{R})=D(\mathbf{R}). \label{Eq.AFPropagator}
\end{eqnarray}
 Further, the quantity $f_\mu(\mathbf{k},\mathbf{q})$ is a form factor describing the
coupling between an electron and a bond pair. For a tight binding lattice and nearest-neighbor bonds,
$f_\mu(\mathbf{k},\mathbf{q})=\cos[(k_\mu-q_\mu)a]$.

The pair correlator of Eq.\eqref{Eq.AFPropagator} can be written in the standard way \cite{PMChaikin},
\begin{eqnarray}
D(\mathbf{R}_m-\mathbf{R}_n)=\langle \tilde{\psi}_m\rangle \langle
\tilde{\psi}^*_n)\rangle+S(\mathbf{R}_m-\mathbf{R}_n)
\label{Eq.UrsellFunction}
\end{eqnarray}
where $\tilde{\psi}_m=\Delta_m\exp{(i\varphi_m)}$ with $\varphi_m=\phi_m$ for
$x$-bonds and $\varphi_m=\phi_m+\pi$ for $y$-bonds (see Fig.\ref{fig.BondLattice})$; S(\mathbf{R})$ is the fluctuation term. In the long-range ordered state below $T_c$, the first term is nonzero. 
In that case, if one neglects effects
of fluctuations i.e. $S(\mathbf{R})$ altogether (as is done in mean-field theory), then one obtains the exact
Gor'kov self energy form \cite{GDMahan} i.e $\Sigma(\mathbf{k},i\nu_n)=\Delta_\mathbf{k}^2/(i\nu_n+\xi_\mathbf{k})$ 
in Eq.\eqref{Eq.StaticSelfEnergy} and corresponding spectral gap $\Delta_\mathbf{k}=(\Delta_d/2)(\cos{k_xa}-\cos{k_ya})$ 
in the N\'{e}el ordered state. Spin-wave-like fluctuations below $T_c$ can be incorporated through $S(\mathbf{R})$ which
generally decays algebraically for large distances i.e. $S(\mathbf{R})\sim R^{-\eta}$ ($\eta>0$, its value
depends on dimension). Above $T_c$, $\langle \psi(\mathbf{R})\rangle=0$ and the only contribution comes from
the fluctuation part. Generically, there is a finite correlation length $\xi$ above $T_c$ and $S(\mathbf{R})\sim \exp{(-R/\xi)}$ or $S(\mathbf{q})\sim 1/[1+(\xi q)^2]$. 

Since we are mainly interested in the spectroscopic features of the pseudogap regime when $T^*(x)$ is
perceptibly higher than $T_c(x)$ so that fluctuations in the pair magnitude $\Delta_m$ are small and short
ranged, we write,
\begin{eqnarray}
D(\mathbf{R})&\simeq& <\Delta(\mathbf{R})><\Delta(\mathbf{0})>\langle
e^{i[\varphi(\mathbf{R})-\varphi(\mathbf{0})]}\rangle \nonumber \\
&\equiv& \bar{\Delta}^2 F(R)
\end{eqnarray} 
where $F(R)=\langle e^{i[\varphi(\mathbf{R})-\varphi(\mathbf{0})]}\rangle$ is the phase correlator.

Analytical expression for the self-energy from Eq.\eqref{Eq.StaticSelfEnergy} can be obtained below $T_c$, 
where quasi-long-range order in purely 2D system or true
long-range order in anisotropic 3D system occurs, as well as above $T_c$ in the temperature 
regime where the exponential decay of correlation is governed by a large correlation length $\xi$
\cite{SBanerjee3}. We have carried out calculations \cite{SBanerjee3} for both anisotropic 3D and 2D cases, while 
incorporating a small interlayer coupling $C_\perp$ (with $C/C_\perp\sim 100$ as suitable for Bi2212) in 
Eq.\eqref{Eq.functional} for the former.
Above $T_c$ the anisotropic 3D system behaves effectively as 2D \cite{PMinnhagen} and our results for various
spectral properties are quantitatively similar and even below $T_c$, for this large anisotropy ratio, qualitative features 
are the same for both the cases. Hence, we
present here the results for the pure 2D system. More specifically, here we have used the form
\begin{eqnarray}
F(R)&=& \left(\tilde{\Lambda}R\right)^{-\eta}e^{-R/\xi} \label{Eq.XYPropagator}
\end{eqnarray}
to calculate the self energy (Eq.\eqref{Eq.StaticSelfEnergy}). Here $\tilde{\Lambda}$ is related to the upper
wave-vector cutoff of the lattice and $\eta=T/(2\pi \rho_s)$ below $T_c$ where $\xi\rightarrow \infty$. Above
$T_c$, we have set $\eta=\eta_\mathrm{BKT}=0.25$.  A combination of MC simulation and well-known
Kosterlitz-Thouless renormalization group relations has been used to estimate $\xi(x,T)$ from the 
functional (Eq.\eqref{Eq.functional}) (see Appendix \ref{appendix.XYModel} for details). The self energy
$\Sigma(\mathbf{k},i\nu_n)$ obtained using the form of $F(R)$ in Eq.\eqref{Eq.XYPropagator} evolves smoothly from 
below $T_c$ (superconducting state) to above $T_c$ (pseudogap state). 

\begin{figure}
\begin{center}
\begin{tabular}{c}
\includegraphics[height=6cm]{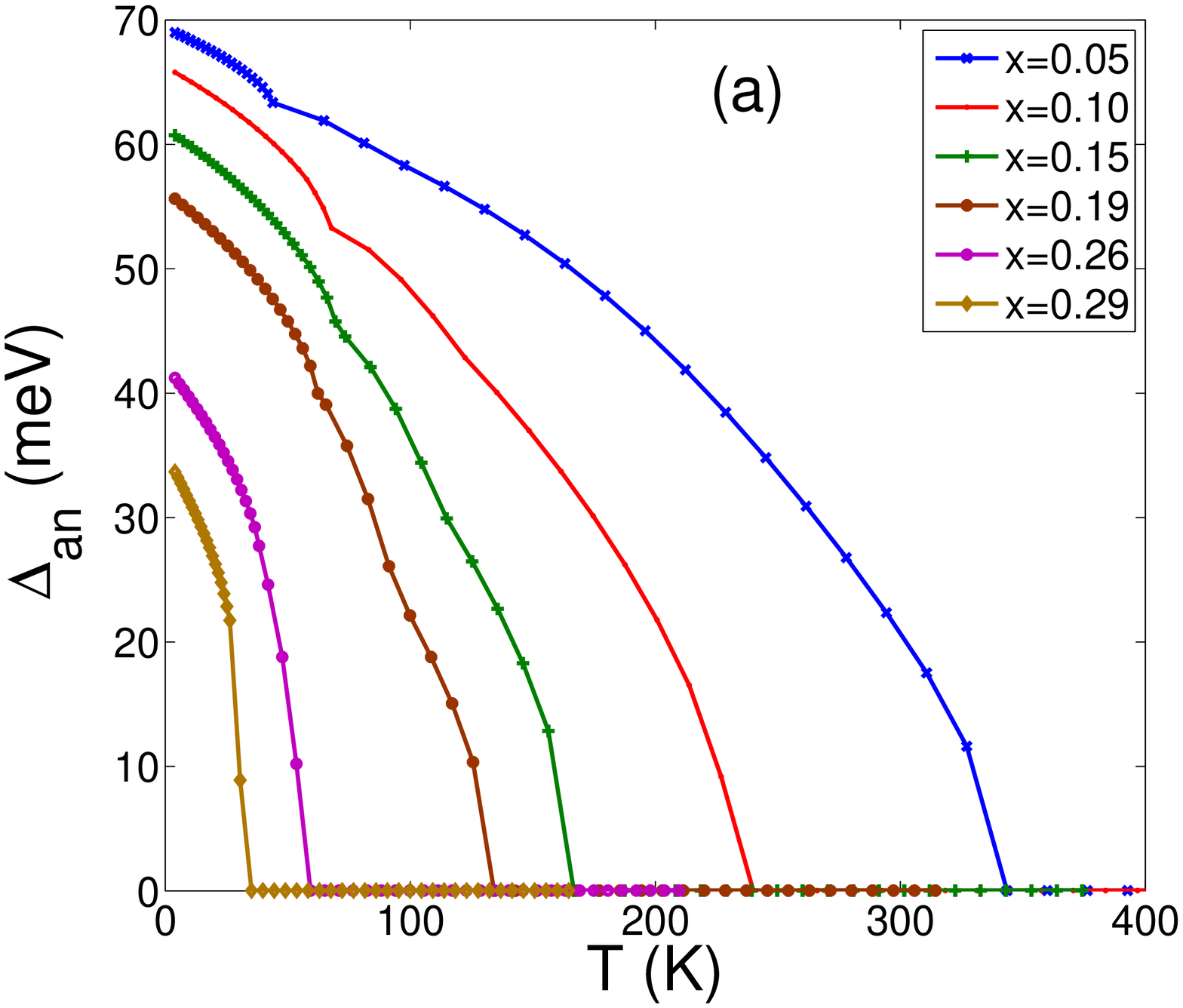}\\
\includegraphics[height=7.5cm,angle=-90]{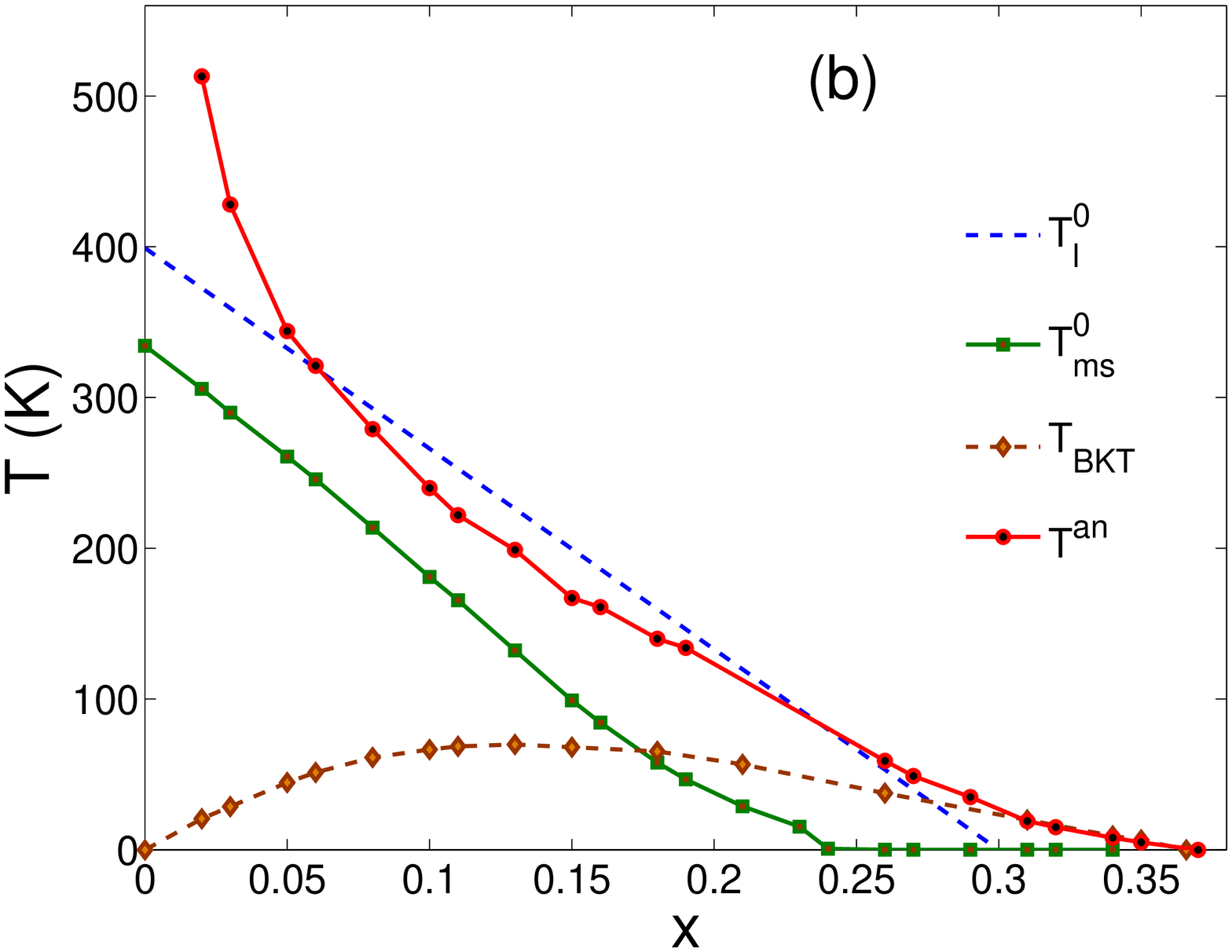}
\end{tabular}
\end{center}
\caption{{\bf (a)} Variation of antinodal gap $\Delta_\mathrm{an}$ with temperature. Slope discontinuities in
$\Delta_\mathrm{an}$ vs. $T$ curves correspond to $T_c$ ($T_\mathrm{BKT}$). {\bf (b)} Pseudogap
temperature scale $T^\mathrm{an}$ obtained from the antinodal gap filling criterion mentioned in the main text.
$T^\mathrm{an}(x)$ is compared with other temperature scales, $T_l^0(x)$, $T_\mathrm{ms}^0(x)$ and
$T_\mathrm{BKT}(x)$. Here, we have taken the nearest-neighbor hopping $t=300$ meV and the next-nearest-neighbor
hopping $t'=-t/4$ \cite{AParamekanti}.}
\label{fig.AntinodalGap}
\end{figure}

For $\mathbf{k}$ on the 
Fermi surface \cite{Footnote4} in the antinodal region, we calculate
$\mathcal{A}(\mathbf{k}=\mathbf{k}_\mathrm{an},\omega)$. Above $T_c$ but below a certain temperature (denoted
as $T^\mathrm{an}$), two peaks appear in $\mathcal{A}(\mathbf{k}_\mathrm{an},\omega)$ at nonzero $\omega$, 
one at $\omega<0$ and another at $\omega>0$, signaling the presence of a pseudogap above $T_c$. The antinodal
gap (denoted as $\Delta_\mathrm{an}$) can be defined from the position of the peak at negative energy
($\omega<0$). This quantity has been plotted in Fig.\ref{fig.AntinodalGap}(a) as a function of temperature for 
a few values of $x$.  

The quantity $\Delta_\mathrm{an}$ goes to zero rather abruptly at $T^\mathrm{an}$, though the average local
gap $\bar{\Delta}$ is non zero above $T^\mathrm{an}$ (see Fig.\ref{fig.Gap}). The antinodal pseudogap fills in at
this temperature \cite{SBanerjee3}. In Fig.\ref{fig.AntinodalGap}(b), $T^\mathrm{an}$ is plotted as a function
of $x$. We notice that this temperature is close to various pseudogap related temperatures e.g. the somewhat
arbitrary linear $T^0_l(x)$ used in Eq.\eqref{Eq.functional}, as well as the temperature scale
$T_\mathrm{ms}^0(x)$ estimated from the temperature dependence of the local gap magnitude. The $x$-dependence of 
$T^\mathrm{an}$ is similar to that of $T^*$ as inferred from ARPES \cite{JCCampuzano1} as well from various
other probes such as Raman spectroscopy \cite{TPDevereaux} and spin susceptibility
\cite{TTimusk,JLTallon} over a rather large range of $x$. 

The picture used in our calculation continues to regard the electrons as coherent at all temperatures whereas there is 
experimental evidence \cite{AIno} that the incoherence temperature is proportional to $x$ so that it is rather small for 
small $x$. Also, for very small $x$, the holes tend to localize, so that a renormalized band theory implying extended 
homogeneous electronic states is inappropriate.

\section{Discussion and Future Prospects}\label{sec.Discussion}
	
We mention here some obvious directions in which the functional and the approach used here need to be
developed. One is to obtain other testable/experimentally measured consequences of the proposed functional.
For example in a magnetic field, the intersite term in Eq.\eqref{Eq.functional} has its phase altered by
Peierls phase factor, as we have mentioned at the end of Section \ref{sec.SpecificHeat}. One should use this to find 
the $T_c(H)$ curve for different values of doping $x$ and thence the `bare' coherence length $\xi_0$ defined through the 
phenomenological equation, $\frac{1}{T_c}\left(\frac{dT_c}{dH}\right)_{T=T_c}=\left(\frac{\xi_0^2}{\Phi_0}\right)$. 
The charge related response of a system described by Eq.\eqref{Eq.functional}, e.g. the
diagonal and off-diagonal components of the conductivity tensor, $\sigma_{xx}(H,T>T_c)$ and
$\sigma_{xy}(H,T>T_c)$, and the Nernst coefficient $\alpha_{xy}(H,T>T_c)$, needs to be calculated and compared
with experiment. Slightly farther afield, the coupling of the field $\psi_m$ to different probes will enable one to 
analyze experimental results obtained e.g. from scanning tunneling spectroscopy, Raman
spectroscopy and neutron scattering. A generalization to a quantum $\psi_m$ functional and inclusion of other
time-dependent effects, e.g. Coulomb interaction and dissipation may enable one to describe quantum
phase-fluctuation effects, which are specially prominent (and decisive) for extreme underdoping \cite{IHetel}.

A very peculiar feature of cuprates is the unusually large proximity effect \cite{YTarutani} observed in them.
While XY-spin-like models have been proposed for this \cite{DMarchand}, a complete understanding of the size,
temperature and doping dependence etc. does not exist. It is possible that the present theory can be
adapted to address this question.

The theory presented needs to be extended in many major ways. For example, there is a lot of experimental
evidence \cite{MAKastner} that the system is a Mott insulator at $x=0$, with a large superexchange $J_{ij}\sim 0.15$ eV., as 
well as for low-energy magnetic correlations in doped cuprates. This antiferromagnetic interaction evolves into 
superconductivity for surprisingly small hole  doping, $x\geq 0.05$. While the crossover and the possibility of coexistence 
have been investigated at $T=0$ \cite{MOgata,TGiamarchi,AHimeda}, there is need for a coupled functional for these two 
bosonic degrees of freedom that goes over to the kind of theory we have described above at large $x$, while it describes an
antiferromagnetic Mott insulator at $x=0$ and persistent spin correlations (including spin density wave
correlations) at $x\neq 0$. Similarly there is considerable evidence for other kinds of
correlations, e.g. nematic \cite{SAKivelson2}, stripes \cite{SAKivelson1}, checkerboard \cite{JEHoffman}, and
charge density wave \cite{CCastellani} whose significance varies with material, doping (including
commensuration effects \cite{ARMoodenbaugh}) and temperature. An appropriate GL like functional is one way of
exploring the details of this competition: attempt in this direction already exist \cite{EDemler}. 
 
The cuprate properties are very sensitive to certain impurities e.g. Zn replacing Cu. Whether this can be
described well in a GL like theory is an interesting question. The effect of impurities or in-plane/intra-plane
disorder is an even more general question in terms of its effect on pairing degrees of freedom as well as
incorporation of this effect in a this kind of picture. A subject of basic interest in cuprate superconductivity is the
possibility of time-reversal symmetry breaking associable with $T^*$ \cite{CMVarma2}. There are at least two
observations, one of Kerr effect \cite{JXia} and another of ferromagnetism with lattice symmetry
\cite{BLeridon}, which seem to point to time reversal symmetry breaking below $T^*$. Since these involve
spontaneous long-range order in circulating electric currents, each within a single unit cell of the lattice,
and these currents can be modeled in a GL functional, one can explore this novel phase and its consequences
in our theory.
 
In conclusion, we believe that the phenomenological theory proposed and developed here not only ties together a range of 
cuprate superconductivity phenomena qualitatively and confronts them quantitatively with experiment, 
but also has the potential to explore meaningfully many other phenomena observed in them.

{\bf Acknowledgments:}  We thank U. Chatterjee for useful discussions. SB would like to acknowledge CSIR (Govt. of
India) for support. TVR acknowledges research support from the DST (Govt. of India) through the Ramanna Fellowship as well as NCBS, Bangalore for hospitality. CD acknowledges support from DST (Govt. of India).  

\appendix

\section{Mean Field Theory} \label{appendix.MFT}

 We describe here various approximate solutions for the properties of the lattice functional
(Eq.\eqref{Eq.functional}). The approximations
discussed here are single-site mean field theory and cluster mean field theory. We also make use of several 
well-known results from the Berezinskii-Kosterlitz-Thouless theory \cite{JMKosterlitz1,JMKosterlitz2,PMChaikin} for XY
spins in two dimensions, in combination with Monte Carlo simulation (see Section \ref{sec.Tc}). For positive $C$ in 
Eq.\eqref{Eq.functional1}, there is a low-temperature phase with long range `AF'
order ($d$-wave superconductivity) or broken symmetry (for $d>2$). The most common
approximation for locating and describing this transition is (single-site) mean field theory, in which we
self-consistently calculate the staggered `magnetic field' $\mathbf{h}=(h_x,h_y)$, acting on the planar spins 
$\mathbf{S}_m=(\Delta_m\cos\phi_m,\Delta_m\sin\phi_m)$, due to its nearest neighbors, assuming it 
to be the same at each site (modulo the sign change due to the two sublattice `AF' order).

In such a mean field theory \cite{PMChaikin,RKPathria}, the self-consistent solution is given by
\begin{eqnarray}
h_\alpha&=&4C\langle S_\alpha\rangle_0 ~~~~(\alpha=x,y) \label{Eq.SelfConsistency}
\end{eqnarray}
with
\begin{eqnarray}
\langle S_\alpha\rangle_0&=&\left(\frac{h_\alpha}{h}\right)\frac{\int_0^\infty \Delta^2 d\Delta P_0(\Delta) I_1(h\Delta /T)}{\int_0^\infty
\Delta d\Delta P_0(\Delta) I_0(h\Delta/T)},
\end{eqnarray}
Here, $P_0(\Delta)=\exp{(-\beta(A\Delta^2+(B/2)\Delta^4))}$ dictates the local distribution (thermal) of gap magnitude, $h=\sqrt{h_x^2+h_y^2}$ is the
magnitude of the `staggered' field and $I_0$, $I_1$ are modified Bessel functions of first kind. 
The transition temperature $T_c$ (which is denoted as $T_c^\mathrm{mf}$ in Fig.\ref{fig.Tc}) satisfies the implicit 
equation
\begin{eqnarray}
2C\langle \Delta^2\rangle_{P_0}|_{T=T_c}=T_c \label{Eq.MFTTc}
\end{eqnarray}
where $\langle \Delta^2 \rangle_{P_0}=\int_0^\infty \Delta^3 d\Delta {P_0}(\Delta)/\int_0^\infty \Delta
d\Delta {P_0}(\Delta)$.
Other physical quantities, such as the superfluid stiffness, the superconducting order parameter, the internal energy
(and its temperature derivative, the specific heat $C_v$), can be obtained using the
self-consistent solution of Eq.\eqref{Eq.SelfConsistency}. For instance, in this approximation, the superfluid density 
$\rho_s$ is given by
\begin{eqnarray}
\rho_s&=&-\frac{C}{2N_b}\langle \sum_{m,\mu} \Delta_m\Delta_{m+\mu}\cos(\phi_m-\phi_{m+\mu})\rangle_0 \nonumber \\
&=&C\sum_{\alpha=x,y} \langle S_\alpha\rangle_0^2, \label{Eq.MFTsfldensity}
\end{eqnarray}

In reality, the field acting on a `spin' fluctuates from site to site. The spatially local fluctuations are systematically included in the well-known cluster theories, the
oldest of which is the Bethe-Peierls approximation \cite{RKPathria}, which consists of a single site coupled to the 
nearest-neighbors which are described by a mean field. We have used it to calculate an `improved'
$T_c$ ($T_c^\mathrm{cmf}$), as shown in the inset of Fig.~\ref{fig.Tc}. 

For small $x$, where amplitude fluctuations can be neglected, an estimate of $T_c$ (denoted as $T_{c,0}$) is obtained by 
replacing $\langle \Delta^2\rangle_0$ in the above relation (Eq.\eqref{Eq.MFTTc}) by $\Delta_{m,0}^2$ that
minimizes the single-site 
term $\mathcal{F}_0$, so that $\Delta_{m,0}^2=-A(x,T)/B$ for $x\leq x_c$ and $\Delta_{m,0}=0$ for $x>x_c$. In this approximation,
\begin{eqnarray}
T_{c,0}&=&\frac{2xc}{2xc+b}\left(1-\frac{x}{x_c}\right)~~~~x\leq x_c \nonumber \\
&=&0~~~~x>x_c \label{Eq.Tc0}
\end{eqnarray}
Here we have neglected the exponential temperature dependence of $A$ (Eq.\eqref{Eq.A}). Consequently $x_\mathrm{opt}$ 
can also be estimated by setting $\frac{\partial T_{c,0}}{\partial x}=0$, which gives
$x_\mathrm{opt}=\frac{1}{2}\left(\sqrt{(b/c)^2+(2b x_c/c)}-(b/c)\right)$. 

If one includes the term $\mathcal{F}_Q$ (Eq.\eqref{Eq.functionalQ}), the self-consistency condition for $T_c$
in Eq.\eqref{Eq.MFTTc} gets modified in the following manner \cite{RFazio},
\begin{eqnarray}
\left(4C\langle \Delta^2\rangle_{P_0}\int_0^\beta d\tau \langle
\cos{\phi_m(\tau)}\cos{\phi_m(0)}\rangle_{\mathcal{F}_Q}\right)_{T=T_c}=1 \nonumber \\\label{Eq.MFTTcQuantum}
\end{eqnarray}
where the average $\langle ...\rangle_{\mathcal{F}_Q}$ is calculated using the eigenstates of $\mathcal{F}_Q$
and the imaginary time on-site phase-phase correlator in Eq.\eqref{Eq.MFTTcQuantum} is given by \cite{RFazio}
\begin{eqnarray}
\langle\cos{\phi_m(\tau)}\cos{\phi_m(0)}\rangle_{\mathcal{F}_Q}=\frac{1}{2}e^{-4\tau V_0(1-\tau/\beta)}.
\end{eqnarray}
where $V_0$ is the on-site Cooper pair interaction strength.

\section{Electron Self Energy in Static Approximation} \label{appendix.SelfEnergy}

The self energy depicted in Fig.\ref{fig.SelfEnergyApproximation} can be written in the following form using
$G_0(-\mathbf{k}+2\mathbf{q},-i\nu_n+iz_m)$ for the internal electron propagator,
\begin{eqnarray}
\Sigma(\mathbf{k},i\nu_n)&=&\frac{T^2}{N}\sum_{\mathbf{q},m} \frac{D(2\mathbf{q},iz_m)
\mathcal{P}(\mathbf{k},\mathbf{q})}{i\nu_n-iz_m+\xi_{\mathbf{k}-2\mathbf{q}}}, \label{Eq.SelfEnergy}
\end{eqnarray}                   
where $D(2\mathbf{q},iz_m)=(1/T)\int_0^\beta d\tau \sum_\mathbf{R} D(\mathbf{R},\tau)
e^{-i2\mathbf{q}.\mathbf{R}+iz_m\tau}$ is the Fourier transform of the time-dependent propagator and
$\mathcal{P}(\mathbf{k},\mathbf{q})=[\cos{(k_xa-q_xa)}-\cos{(k_ya-q_ya)}]^2$. If the pairs acquire a finite 
lifetime $\tau_p$, the pair correlator can be represented in terms of the product of the static propagator
(Eq.\eqref{Eq.AFPropagator}) and a time-dependent part as $D(\mathbf{R},t)=D(\mathbf{R})e^{-t/\tau_p}$ so that
$D(2\mathbf{q},iz_m)=(1/T)(e^{i\beta/\tau_p}-1)D(2\mathbf{q})/(iz_m+i/\tau_p)$. This form indicates that pair correlations 
decay temporally with a lifetime $\tau_p$ (one can instead take an oscillatory form i.e. $D(\mathbf{R},t)\sim
\cos{(t/\tau_p)}$ but this does not change our main conclusion). One can perform the summation over the bosonic
Matsubara frequencies ($z_m$) in Eq.\eqref{Eq.SelfEnergy} with the aforementioned form of
$D(2\mathbf{q},iz_m)$ and obtain
\begin{eqnarray}
&&\Sigma(\mathbf{k},iz_m) \nonumber \\
&&=\frac{1}{T}\sum_\mathbf{q}\frac{D(2\mathbf{q})\mathcal{P}(\mathbf{k},\mathbf{q})
((1-e^{i\beta/\tau_p})f(\xi_{\mathbf{k}-2\mathbf{q}})+e^{i\beta/\tau_p})}{i(\nu_n+1/\tau_p)+\xi_{\mathbf{k}-2\mathbf{q}}}.\nonumber \\
&&
\end{eqnarray} 
Here $f(\omega)=1/(e^{\beta\omega}+1)$ is the Fermi function. When $T>>(1/\tau_p)$ (also $\nu_n>>(1/\tau_p)$
since $\nu_n\propto T$) i.e. inverse pair lifetime is much smaller than $T$, the self energy given above would
effectively reduce to the form given in Eq.\eqref{Eq.StaticSelfEnergy}.
\section{Estimation of Correlation Length $\xi$} \label{appendix.XYModel}
 We estimate $\eta=T/(2\pi \rho_s)$ (below $T_c$) and $\xi$ (above $T_c$) that appear in
Eq.\eqref{Eq.XYPropagator}. As already discussed, we calculated $\rho_s$ below $T_c$ from our functional in 
Section \ref{sec.SuperfluidDensity} by performing MC simulation. Correlation length $\xi$ can be
estimated by fitting obtained $\rho_s(x,T)$ below $T_c$ with the BKT form, 
$\rho_s(x,T)=\rho_s[T_c^-(x)][1+b(x) \sqrt{T_c(x)-T}]$ with $\rho_s(T^-_c)/T^-_c=2/\pi$, and $b(x)$ and $T_c(x)$ as 
fitting parameters. BKT RG relates \cite{VAmbegaokar} $b(x)$ to the temperature-dependence of $\xi$ above
$T_c$ through $\xi(x,T)\simeq a_0 \exp{\left[b'(x)/\sqrt{T-T_c(x)}\right]}$, where $bb'=\pi/2$ and $a_0$ is a microscopic 
length scale of the order of the lattice spacing.


\begin{thebibliography}{10}
\bibitem{Footnote4} The Fermi surface has been defined from the locus of $\mathbf{k}$ points for which
$\xi_\mathbf{k}=0$ in the Brillouin zone. We have done calculations using other criteria of determining Fermi surface 
e.g. from the locus of the maximum of $\mathcal{A}(\mathbf{k},\omega=0)$. For these different criteria as well, 
the main features of our results remain unaltered with only slight modifications in the details. 
The chemical potential $\mu$ is calculated by setting
$\int_{-\infty}^\infty d\omega\sum_\mathbf{k}f(\omega)\mathcal{A}(\mathbf{k},\omega)=1-x$
($f(\omega)=1/(\exp(\beta \omega)+1)$ is the Fermi function).
\end{thebibliography}
\end{document}